\documentclass[11 pt, letter paper]{article}
\usepackage[margin= 0.9 in]{geometry}
\usepackage{epigraph}
\usepackage{amsmath,amsfonts,amssymb,setspace}
\usepackage{wrapfig}
\usepackage{tikz,graphicx,xcolor}
\usetikzlibrary{calc,arrows,babel}
\usepackage{physics,xparse,mathrsfs}
\usepackage{hyperref}
\hypersetup{
colorlinks=true,
linkcolor=blue,
filecolor=magenta,
urlcolor=blue,
}
\spacing{0.93}
\title{On the Influence of Gravity on the Propagation of Light\\ Some Comments on the A. Einstein's \(1911\) Paper\\Light Interaction With Gravity\\(The Mistake Made In $1911$)\\ \(1911 - 2022\)}
\author{By \\Rigoberto Martinez\\Physicist}
\date{July \(31\), \(2022\)}

\begin{document}
\maketitle
\begin{abstract}
This is a review with some comments on the A. Einstein's \(1911\) paper, which he published as one of his many attempts of the general theory of relativity. The main point of the idea is to propose a new approach about light and its motion  as well as study the assumption made by A. Einstein concerning the possibility of the variation of the speed of light in presence of Gravitational Field.\\[3 pt]
\end{abstract}
\section*{\S$1.$ \,A. Einstein's \(1911\) paper\\ \center First Part}
When A. Einstein realized, in \(1907\), the incompatibility of his ideas with the Newtonian's concepts about  gravity he tried to reconcile his  ideas  with the special relativity and find out a new form to treat gravity using the equivalence principle as a light. Even when H. Minkowski found the geometry meaning of the Lorentz transformation A. Einstein did not use it at all in \(1911\).\\[3 pt]

In his article he derived some physical phenomena using the equivalence principle and the propagation of light in a fixed gravitation field. All these consequences were obtained by A. Einstein and showed the importance of the idea between local gravity and non-inertial reference systems. There is no doubt that all these consequences obtained were very unusual during that time, however could there be  another form to obtain the same results in particular in relation with the bent and propagation of light in the presence of gravitational field.

The idea in \(1911\) that was used to derive the bent of light is based on the assumption that the speed of light could change in a presence of gravitational field, is say; the speed is a function of the points in the presence of gravity as Einstein deduced: 
$$c=c_0 \left( 1+ \frac{\phi}{c^2} \right)$$

The equation above allowed A. Einstein to derive the angle of deflection of a light when passes  through of gravitational field produces by a heavenly body for instance the Sun. The result obtained was an half of the experimental got in \(1916-1919\) which was predicted using all the frame-work of the general theory of relativity. Go back in \(1911\) the assumption made about the possibility of the change of the speed of light expresses by the equation deduced by Einstein is a heuristic one, from my point of view, to face the problem there is a nice way that possibly one can obtain the same results and all the physical consequences deduced from it as well.\\[3 pt]

Let's now recap to refresh the memory about the assumptions made by A. Einstein in \(1911\) as we could think were used at that time, this I supposed were made by him and expressed it in his \(1911\)paper.
\begin{enumerate}
\item The Equivalence Principle (Local gravity phenomenon could be mimic as an effects of no-inertial reference systems). 
\item The propagation of light in a presence of gravitational field could be treated as wave phenomenon.
\item Static gravitational field (At that moment of time there were not relation between the gravitational potential and the metric tensor \(g_{\mu \nu} \) even though H. Minkowski had been published his memory unifying space and time but A. Einstein did not believe that was useful for his seeking).
\end{enumerate}
After being mentioned the assumptions  made, is time to analyze what Einstein proposed in \(1911\) in what is the comments about this issue. In order to do so is necessary to explore how light propagate, so as to obtain the same result as Einstein got.
\section*{\S$2.$ \,Propagation of Light}
In his article Einstein\cite{einstein1911influence} assumed that the propagation of light through space and in a presence of gravitational field is as a wave phenomenon, so by means of this idea light could be undergo deflection and this could be study using Huyghen's principle as can be read in his \(1911\) paper where was obtained a first approximation the deflection or the expected angle deviation \( \delta \alpha\). However the same result was obtained many years before him, in \(1804\) with the idea that really light could propagate as point-like particle with a constant speed of \( c\) and using Newton's ideas of gravity as Einstein did as well.\\[3 pt]

So the two approaches gave the same value for the angle deflection \( \delta \alpha = 0.875\) seconds of arc. Even though Einstein's approach was in relation of his idea of Equivalent Principle seems more natural the Newtonian's approach even when the treatment gave the same result, however there is nothing new in that. Why both approaches gave the same result? is something interesting to think of. Now taking into account Einstein's point of view of the propagation of light one could ask, Why this result can not be obtained from the Electromagnetism frame-work? if light is not a gravitational phenomenon and if that is possible, which I think it is. It is necessary to modify the general theory of gravity or both but we are getting sidetracked, so let's continue with the discussion.\\[3 pt]

Is amazing to think that A. Einstein even after having developed his theory pointed out the idea that the speed of light could be dependence on the gravitational phenomenon and which is even more enchanting is that, all the electromagnetism phenomenon does not appear in the general theory and some properties was covered only by the special theory of relativity. The only link as many specialist point out even Einstein did is through the Energy-momentum Tensor, \(T_{\mu \nu}\). For instance, can be read in \cite{whitrow1950structure} some of the mentioned in the paragraph, which I encourage to the reader to do so.\\[3 pt]

As Einstein thought in \(1911\) there is no need to consider Maxwell`s equation in order to calculate the deflection angle and study light propagation in space, just assume that light is an EM wave phenomenon, even in a presence of gravitational field and from that the only way light can bending (A natural way due to the assertion made then) is under the assumption of the changed of the speed of light given by the equation: \[ c=c_0 \left(1+ \frac{\phi}{c^2} \right) \] 

Of course the angle deflection is a first order of magnitude, not consider \(0(c^4)\) as Einstein did in his calculations of the \(1911\) paper. However the result of bending light can be obtain without Einstein`s idea of light speed variation. Is also interesting that this idea of possible change in the speed of light is not used to derive the other results such as the change of the frequency of light in homogeneous and static gravitational field.\\[3 pt]

\section*{\S$3.$ \, Energy Conservation Law and Light Propagation}

Instead of consider a wave light propagation, let's use the quantum picture of light and parallel to this make use of the energy conservation applied to conservative system like we are going to study. Let's imagine a process of interaction between light and the gravitational field in small region of flat space-time where the gravity potential is given and fixed and try to get the same result as Einstein got. By doing use of this idea it is possible to write the following equation:
\begin{equation} \label{equ1}
\hbar \omega+\frac{\hbar \omega}{c^2} \phi=\hbar \omega'+\frac{\hbar \omega'}{c^2}\phi' 
\end{equation}

From the equation [\ref{equ1}] and taking in consideration the equivalence principle, we can reduce the region of study in such manner that the equation [\ref{equ1}] could be written as follows:
\begin{equation}\label{equ2}
-\Delta \omega= \frac{\phi \Delta \omega}{c^2}+\frac{\Delta \omega \Delta \phi}{c^2}+\frac{\omega \Delta \phi}{c^2}
\end{equation}

Keeping in mind , in equation [\ref{equ2}] that we are study small region with a given gravitational field is possible to get:
\begin{align}   \label{equ3}
- \left(1+ \frac{\phi}{c^2} \right) \mathrm{d}\omega &=\frac{\omega \mathrm{d}{\phi}}{c^2}\\
   \omega &= \frac{\gamma}{\left(1+\dfrac{\phi}{c^2} \right)}\label{equ4}
\end{align}

The equation [\ref{equ4}] represents the dependence of the frequency of the photon as a function of the gravitational potential at a point \(x\). From this equation is possible to obtain the significance of the constant \(\gamma\), if we consider the motion of the photon far away of any source of gravity, is say; when \( r \rightarrow + \infty \) then \(\phi \rightarrow 0\). Therefore is possible to write the following equation which is one of the many results obtained by A. Einstein in first approximation:
\begin{align} \label{equ5}
 \omega &= \dfrac{\omega_0}{ \left( 1+ \dfrac{\phi}{c^2}\right)}
\end{align}

The equation [\ref{equ5}] tell us that far away from any source of gravity the photon carry energy of the amount of  \(E= \hbar \omega_0\),which can not be take it out. At first glance, only this result is something to put attention on it. \footnote{Bear in mind that this statement is new, photon could interact with the gravitational field in sense of the Compton Effect  but in a more complicated way}  All this suggest that really the process is an interaction between the gravitational field and the corpuscular side of the electromagnetism phenomenon,photon, and is not a simple propagation of EM wave in the region which there is gravitational field. To be honest is important to give more thought on this issue and get a final and sound argument on how really light propagate in a presence of gravity but at this point it seems sensible that the conclusion written above is possibly solid. A study of the propagation is crucial in order to comprehend really what is the issue here with electromagnetism and gravity.\\[3 pt]

\section*{\S$4.$ \, Light and the Frequency Shift}

In vacuum, light propagate with a constant speed in all possible directions. That is the statement said by A. Einstein in \(1905\) which was used to build up his theory of special relativity, with experimental support, However in \(1911\) he changed his mind about it. Even in his general theory of relativity pointed out that in a strong gravitational field this postulate will not be more valid and the speed should change under those circumstances.\\[3 pt]

In order to continue the main idea, is reasonable to postulate the following statement and from it we can deduce all possible physical consequences, so let's establish that: \\[3 pt]

\texttt{ Let's consider that in a weak gravitational field the speed of light should remains constant as well as in vacuum}. Following this idea is possible to write an equation which could be consider as identity, some kind of constrain, and a fundamental as well, so we have:
\begin{align*}
c &=\nu_1 \lambda_1=\nu_2 \lambda_2 = \text{constant}\\
\omega_1 \lambda_1 &= \omega_2 \lambda_2
\end{align*}

In light of this we can write the two main equations  need it to derive some physical conclusions as follows, also considering at the same time equation [\ref{equ4}].
\begin{equation} \label{equ6}
\left \{
  \begin{aligned}
  \omega_1 \lambda_1 &=\omega_2 \lambda_2= \cdots\\
  \omega &= \dfrac{\gamma}{\left(1+\dfrac{\phi}{c^2}\right)}\\
  \arrowvert \phi \arrowvert & >0
  \end{aligned}
  \right.
\end{equation}
With this, we can obtain some physical results by playing with those equations [\ref{equ6}]. The third equation is just the condition that gravitational potential is negative due to the fact that we make at $+\infty$ the gravitational potential equal to $0$. Consider a small region and taking valid, of course, the equivalence principle  we are going to try to derive the frequency shift of light using [\ref{equ6}].\\[3 pt]

Imagine two points in a given gravitational field $\phi$. Let at point $x_2$ be the potential $\phi=0$ and let at point $x_1$ be the potential $\phi$ both points lies on the same potential line. Using the conservative property of gravitational field and using the equations [\ref{equ6}], so we have:

\begin{align*}
\frac{\gamma \lambda_2}{\left(1+\dfrac{0}{c^2}\right)}&=\frac{\gamma \lambda_1}{\left(1+\dfrac{\phi}{c^2}\right)} \\ \text{which can be reduced as follow:}\\
\lambda_2 \left(1+ \dfrac{\phi}{c^2} \right)&=\lambda_1
\end{align*}

After some algebraic manipulations and having been done all the intermediate mathematical steps, we can get the follows relationship between $\lambda$ and the potential gravitational field \(\phi\):
\begin{equation} \label{equ7}
\begin{aligned}
\lambda_2-\lambda_1 &=- \dfrac{\phi}{c^2} \lambda_2\\ \text{ Considering the third equation in {[\ref{equ6}] ,we have:}}\\
\dfrac{\lambda_2-\lambda_1}{\lambda_2}&=-\dfrac{\phi}{c^2}>0 \leftrightarrow \lambda_2>\lambda_1 \leftrightarrow \nu_1>\nu_2\\
\dfrac{\Delta \lambda}{\lambda_2}&=-\dfrac{\phi}{c^2}\\
\end{aligned}
\end{equation}
The last equation in [\ref{equ7}] is the well-known relationship between the wave-length of the EM, photon, with the gravitational potential. Which is the same mathematical expression A. Einstein got in his $1911$ paper. Even though here in this analysis appear the wave-length instead of the frequency, it is easy to demonstrate that both expressions are completely equivalents. So we get the same expression using a different approach but in the end the physical meaning and processes are basically identical, both goes one to one hand.\\[3 pt]

\begin{equation} \label{equ8}
\dfrac{\Delta \lambda}{\lambda_2}=-\dfrac{\phi}{c^2}
\end{equation}

The equation [\ref{equ8}] shows us the frequency shift of the Electromagnetic phenomenon, or using the corpuscular side; the loss or gain in the energy of the photon in a gravitational potential field, either way upward or downward motion.\\[3 pt]

Considering the general case of an atom, for instance in the Sun which emitted a photon and it is received on the Earth. We have in this particular case that the gravitational potential at these points are $\phi_1\,\, ,\, \phi_2$ and using again the equations $(1,2)$ in [\ref{equ6}], we have:

\begin{equation}\label{equ9}
   \begin{aligned}
\dfrac{\gamma \lambda_1}{\left(1+ \dfrac{\phi_1}{c^2}\right)}  =\dfrac{\gamma \lambda_2}{\left(1+ \dfrac{\phi_2}{c^2}\right)}\\ \text{which is equal to:}\\
\lambda_2=\left(\dfrac{ 1+\dfrac{\phi_2}{c^2}}{ 1+\dfrac{\phi_1}{c^2}} \right) \lambda_1
   \end{aligned}
\end{equation}

The above equation [\ref{equ9}] can be rearrange as follows and could be compared with the equation in the $ 1911$ paper, which is basically the same expression with the same physical meaning(have been obtained the same result):
\begin{align} \label{equ10}
\dfrac{\lambda_2-\lambda_1}{\lambda_1}=\dfrac{\left( \dfrac{\phi_1-\phi_2}{c^2}\right)}{\left(1+\dfrac{\phi_1}{c^2}\right)}= \dfrac{\Delta \lambda}{\lambda_1}
\end{align}

If in the mathematical expression [\ref{equ10}] we consider that the gravitational potential could be a small quantity, is possible to make the following approximation which results coincides with the limit case found in the general theory of gravity, is say: \\[3 pt]

\begin{align*}
\dfrac{\Delta \lambda}{\lambda_1}\approx \left( \dfrac{\phi_1-\phi_2}{c^2} \right)
\end{align*} 

Throughout the paper I don not have mentioned anything about the observer which play a crucial role in the paper published by A. Einstein, however the importance of observers is implicit since to compute all the calculations is essential to know where the radiation is coming from, in other words; which point in a flat space-time is emitting or receiving the photon, EM, so with that information we can get from it a much more valuable physical information.\\[3 pt]

For instance, let's imagine that the point called $x_2$ is on the surface of the Earth and consider also a homogeneous gravitational field as always. Let at $x_1$ be the point at which the potential is equal to $\phi$ so in this case the equation [\ref{equ8}], the right-hand side is negative so we can write it in the form:
\begin{equation} \label{equ11}
\begin{aligned}
\lambda_2-\lambda_1=-\dfrac{\phi}{c^2} \lambda_2\\ \text{consider the relation:}\\
\lambda_1=\dfrac{c}{\nu_1} \,\,\, ,\,\,
 \lambda_2=\dfrac{c}{\nu_2}\\
 \nu_1-\nu_2=-\dfrac{\phi}{c^2}\nu_1
\end{aligned}
\end{equation}

From equation [\ref{equ11}] we can obtain the well-known result from the special theory, is say:

\begin{align*}
\dfrac{\nu_2-\nu_1}{\nu_1}=\dfrac{\phi}{c^2}
\end{align*}

Whereby a shift of the spectral lines of the source of light would be measured at point $x_1$. All the comments is nothing new, however is surprisingly interesting that using the conservation law of energy and the \texttt{assumption of the constancy of the speed of light in weak gravitational field led us}. Other physical phenomena can be studied as well, using either Einstein's point of view or the all done above. Which could be sensible to imagine is that all this can be found using the general theory of gravity or the fact that by means which gravity and EM field interact is through an intrinsic and complicated fields interaction process rather than just making a conjecture that light propagate as a wave or using the corpuscular side, the quantum aspect.\\[3 pt]

I will be eager to face this task perhaps in another article. Meanwhile let's continue with the task of what we are trying to do. All the statement found by Einstein is correct in first order of magnitude as he mentioned in his article. In this approach there is none of that kind of approximation, perhaps will be appear in the computation of the expected deviation angle, but this does not mean that all what the reader could find here is a replacement of Einstein's ideas it's rather a support of his point of view in Physics.\\[3 pt]

\section*{\S$5.$ \,\, Deflection of Light in Flat Space-Time}

Let's calculate first if a light-ray could be trapped by a heavenly body under the studied conditions. Imagine a photon that moves from a region with no gravity presence and it approaches to a heavenly body of mass $M$ from the equations [\ref{equ6}] we can have the total energy of the conservative system as:
\begin{equation} \label{equ12}
\begin{aligned}
\mathcal{E}&=\hbar \omega-\dfrac{\hbar \omega}{c^2} \phi\\ \omega&=\dfrac{\omega_0}{\left(1+\dfrac{\phi}{c^2}\right)}
\end{aligned}
\end{equation}

Substituting the second equation of [\ref{equ12}] into the first and avoided some steps, we have:

\begin{equation} \label{equ13}
\begin{aligned}
\mathcal{E}=\hbar \omega_0 \left(\dfrac{1-\dfrac{\phi}{c^2}}{1-\dfrac{\phi}{c^2}}\right)=\hbar \omega_0 \\
\mathcal{E}=\hbar \omega_0= \text{Constant}>0
\end{aligned}
\end{equation}

Which clearly means that the light-ray can not be trapped by any gravitational field under the assumption made in this analysis. Since from celestial classical mechanics is well-known that only when the total energy of the system is \(\mathcal{E}<0\) should be exists a closed trajectory either a circle or ellipse, therefore by this perspective is possible for light only a small deviation from its initial path. To consider the possibility that light could be trapped by a gravitational field is necessary other considerations coming from like general theory of relativity, electromagnetism or even quantum mechanics.\\[3 pt]

Is understandable to see that the energy condition \( \mathcal{E} \geq 0\) permits only a parabolic or hyperbolic trajectory, However between these two options the only one reasonable it is the hyperbolic trajectory due to  \(\mathcal{E}>0\) and from this we can deduce an expect small angle of deviation \(\delta \alpha\). In other words from the analysis done above we arrived to the sound conclusion that light would be undergo a deviation from its original path in a region where there is a weak gravitational field and goes beyond this we could say that in \texttt{any gravitational field light undergo deflection and its speed, if I do not mistaken, should remains constant} which is a counter argument about what A. Einstein wrote in his \(1911\) paper and some comments he expressed implicitly in the successive developments of his general theory of gravity  \cite{whitrow1950structure}. 
\section*{\S$6.$ \, Expected Angle of Deviation}
From equation [\ref{equ13}], we know beforehand the possible trajectory that light could take when passes through a heavenly body like the sun, however try to obtain the expected angle deflection is not so easy. So in order to do that, we need to take some concepts from the geometry point of propagation of light as tenable argument, \(\lambda \rightarrow 0\), under this condition and from equation [\ref{equ6}] we can use the eikonal equation to attempt in first  approximation \(\delta \alpha\).\\[3 pt]

As is now from geometry light theory there is a relation between these physical  magnitudes \(\psi\) and \(\omega\), is say:
\begin{equation} \label{equ14}
\begin{aligned}
 \dfrac{\partial{\vec{k}}}{\partial{t}} &=- \dfrac{\partial \omega}{\partial {\vec{r}}}
\ \text{And the equation}\\
 \omega&= \dfrac{\omega_0}{\left(1- \dfrac{\phi}{c^2}\right)} \approx \omega_0 \left(1+\dfrac{\phi}{c^2} \right)
 \end{aligned}
\end{equation}
Taking the partial derivative we have:
\begin{equation} \label{equ15}
\dfrac{\partial\omega}{\partial{\vec{r}}}= \vec{\nabla} \omega=-\dfrac{\omega_0 GM}{c^2}\dfrac{\hat{r}}{r^2} 
\end{equation}
With equations [\ref{equ14}] and [\ref{equ15}] we can deduce the variation in the \(\Delta \vec{k}\) very easily, so we get:

\begin{equation}\label{equ16}
\begin{aligned}
-\hbar \Delta \vec{k} \cdot \hat{r}&=-\dfrac{GM \hbar \omega_0}{c^2} \int \dfrac{\mathrm{d}s}{r^2}\\
\hbar\arrowvert \Delta \vec{k} \arrowvert=\dfrac{GM\hbar \omega_0}{c^2 \Delta} \int \limits_0^\pi \sin\phi \mathrm{d}\phi&=2\dfrac{GM\hbar \omega_0}{c^2 \Delta}\\
\hbar \arrowvert \Delta \vec{k}\arrowvert&=2 \dfrac{\hbar \omega_0}{c} \sin \left( \dfrac{\delta}{2} \right)
\end{aligned}\\
\end{equation}

Equating the last two equations of [\ref{equ16}], we can get in a first approximation the expected angle as A. Einstein got in his \(1911\) paper, with the consideration of weak gravitational field but maintaining constant the speed of light in presence in such kind of field. I suggest, that this proposal must be a necessary request to Nature in the light that we obtained the same results without the Einstein's conjecture of variable speed of light as he suggested in his earlier works on general theory of gravity, which have to be called a law of general field geometry gravity. finally we get:\\[3 pt]
\begin{equation*}
\begin{aligned}
 \sin \left( \dfrac{\delta}{2}\right)= \dfrac{GM}{c^2 \Delta} \Leftrightarrow \delta \approx 2 \dfrac{GM}{c^2\Delta}
\end{aligned}
\end{equation*}

A full derivation from the general relativity of the expected angle deviation can be found in \cite{mcmahon2006relativity} which take into account  Einstein's idea on gravity,  is say, gravity as geometry model where the potential gravitational are encoded in the metric tensor: \( g_{\mu \nu}\) or in \cite{matveev1976mechanics} using Newton's ideas and consider light as flow of particles. In the limit case of weak gravitational field this classical potential are the temporal component of the metric:\[g_{00}=1+2\dfrac{\phi}{c^2},\, \, \arrowvert \phi \arrowvert>0\] of course in first order of approximation. With this the line element \footnote{Could be interesting to derive this result using the main idea developed here as interaction process between the photon and the gravity phenomenon} of a non-zero mass particle could be written as:

$$  \mathrm{d}s^2 \approx\left(1+2\dfrac{\phi}{c^2} \right)\mathrm{d}x^{02}-\mathrm{d}\vec{r}^2 $$

Which the speed of light is a fundamental constant even in the presence of a weak gravitational field. However in the \(1911\) paper there was not such idea and basically all derivation was done without have been used this idea but in order to obtain the expected angle deviation a variation speed was proposed by him whereby permits him to treat the light propagation as a wave and possible bending was calculated but in first order of magnitude and in unnatural way to got it,  neglecting terms like $\order{c^4}$. 
\section*{\S$7.$ \, What Does all this mean?\\ A kind of conclusions of This First Work}

First of all, there is nothing new in this paper, apart for the idea of interaction process mentioned many times above, photon-gravity, which are not known today for the physicist community around the world however the main points was to exposure the Einstein's ideas in \(1911\) concerning the implicit postulate that speed of light must change in presence of gravitational field. Under this light was possible to compute an expected angle deviation using the wave character of light. In counter of this argument which seems reasonable, it was demonstrated such a variation of the speed is not a sound argument and is unnecessary in calculate the deviation angle of the light from its original path.\\[3 pt]

I am not contradicting what Einstein did in \(1911\) but certainly was not careful in that aspect concerning of the propagation of light as wave, I comprehend that if light propagate as wave the possible mechanism for an expected angle deviation was using the notions coming from wave aspect of the light. Moreover we can found that the same value of the expected deviation angle was computed using the idea of corpuscular aspect and the value was practically the same in order of magnitude. Besides all this, Einstein computation of the angle deviation appears not natural from the physical point of view. Let me put again the equation which is easy to see the major problem with the concept of variation and the necessity of taking only the first order of magnitude and neglect terms of \(O(c^4)\);
\begin{equation} \label{equ17}
c=c_0 \left(1+ \dfrac{\phi}{c^2} \right)
\end{equation}

As is easy to see there is not use of electromagnetism theory to predict the deviation angle even in his general theory of relativity is not possible to obtain it using both electromagnetism and gravity, not to consider the \(T_{ \mu \nu}\), it is just necessary use gravity as a physical field phenomenon . All the aspects of light in general relativity is absence and only in the special theory some properties are explains and incorporated. Therefore is something to take into consideration and perhaps an indication of the incompleteness of the general theory of relativity.\\ [3 pt]

So speed of light should remains constant in a weak gravitational field and perhaps in a strong gravity \texttt{have to remains constant}. The reason for what this is no so clear to me now is in relation to the face of many mathematical walls however a possible geometrical constrain could gives us an answer to this crucial issue. I comprehend the question is difficult and much more to figure out a possible solution or near solution but if were possible the unification between gravity and electromagnetism a solution will coming from it.\\[3 pt]

As is known \( F_{\mu \nu}\) remains unchanged under the presence of gravity, which means that there is not interaction between these two fields that dominate  basically the macro-world. After all this, the assumption made here about left constant the speed of light was successful, because the same physical phenomena were obtains and even the expected deviation angle. Which indicates that not only in vacuum the speed is a constant, also and the presence of gravity should remains constant. Doing this we avoided some major difficulties in deal with the equation [\ref{equ19}]. I am not interesting in go beyond what I did, for now, but I hope this will help others to think about what are the main difficulties physics nowadays is facing.\\[3 pt]
\vspace{-1 cm}
\begin{center}
Submitted date: August $22$, $2022$\\
First Paper
\end{center}
\vspace{-1 cm}
\begin{center}
\begin{tikzpicture}
\draw (3,0) -- (11,0);
\end{tikzpicture}
\end{center}
\section*{\S $8.$ \, Energy and Momentum of Light\\Its Minimum Energy Value.}

Is very known since $1900-1905$ is possible to associate to light both energy which is a function of its frequency $\mathcal{E}=h\nu$ and  momentum $\mathcal{E}=pc$. The second equation is consequence of the fact that rest-mass of the photon is zero so basically is one of many results from the special theory of relativity, is say, come from the relativistic equation between energy and momentum $ \mathcal{E}^2=p^2c^2+m_0^2c^4$. What perhaps kind of new is what was mentioned in the article \cite{martinez2022influence} about a minimum value concerning the energy of the photon far away from gravitational field or others kind of interactions, classically speaking of course.\\[3 pt]

This minimum energy value it will be denoted as:
\begin{equation}\label{equ18}
 \begin{aligned}
 \omega= \dfrac{\omega_0}{\left(1+\dfrac{\phi}{c^2}\right)}\,\, \abs{\phi}>0\\
 \text{Where now:}\\
 \mathcal{E} =\hbar \omega_0=h \nu_{\text{min}}, \, \, \phi \rightarrow 0 \,\,\text{when} \,\,r\rightarrow \infty 
 \end{aligned}
\end{equation}

So the idea is that this value frequency $\nu_{\text{min}}$ can not be taken it out from the photon, in other words the energy associated with it is the minimum value possible that the photon could has, isolated from any kind of interactions. It can be summarize saying that the photon can not loss all its energy, so the limit value is: $ \mathcal{E}=h\nu_{\text{min}}$. For simplicity it will noted as: ${\mathcal{E}}_0=h{\nu}_0$. Concerning how possible is to obtain this value experimentally is a difficult issue to think of. In the other hand, we could possible think about its consequences for instance in the cosmology aspects. How the expansion of the universe could affects this energy value of the photon? Would exists some kind of constraints due to this energy value and could affect to the expansion of the Universe? This point turns out to be interesting.\\[3 pt]

Assuming the two relations related with light we can deduce the Doppler effect with the aid of the Lorentz energy transformation:
\begin{equation}\label{equ19}
\begin{aligned}
\mathcal{E}=\gamma \left( \mathcal{E^\prime}+\beta c{p\prime} \right)\\
\text{Putting the follows identity}\\
\mathcal{E^\prime}=h{\nu^\prime}, \, \, \mathcal{E}=h \nu \, \text{and} \,\; p=\dfrac{\mathcal{E^\prime}}{c}\\
\text{We have}\\
\nu=\gamma \left( \nu^{\prime}+\beta v^{\prime} \right) 
\end{aligned} 
\end{equation}

From equation [\ref{equ19}] we get the well-known result:

\begin{equation}\label{equ20}
\nu^\prime=\nu\; \sqrt{\dfrac{1-\beta}{1+\beta}}
\end{equation}

Which is nothing new but the derivation is much easier than others. What seem to me interesting is the used of the idea of the quantum side of the electromagnetic phenomenon. Having in mind that what was obtained above is just the classical Doppler effect which are very well studied. A second derivation is given by the follows procedures:\\

If we imagine  two reference system being one of them a stationary observer whose emission source of light is located and the other observer is moving with a constant speed along the$x$-axis, we can compute the relation between these measurements of the two wavelengths by the following equation:\\
\begin{equation}\label{equ21}
\begin{aligned}
\dfrac{1}{\lambda} &=\gamma \left(\dfrac{1}{\lambda^{\prime}}+\dfrac{\beta}{c}\nu^{\prime}\right)\\
\lambda \nu=\lambda^{\prime}\nu^{\prime}&=\text{Constant}
\end{aligned}
\end{equation}

  Using the second equation [\ref{equ21}] as a constrain, we can obtain:
  \begin{equation}\label{equ22}
  \begin{aligned}
  \nu^{\prime} \gamma \left(1+\dfrac{u}{c}\right)\lambda=\lambda\nu\\
  \beta=\dfrac{u}{c}\\ \text{We have}\\
  \nu^{\prime}=\nu\, \sqrt{\dfrac{1-\beta}{1+\beta}} \quad u>0, u<0\\
  \end{aligned}
  \end{equation}
  
  From this point of view as always happen in any physical model, there is a kind of problem so in this case there is a big problem when the velocity $u$ tends to $c$ is say, $u \rightarrow \, c$. Therefore as consequence of this, the energy can goes up at any value or goes down up to be zero and those kind of thing are very problematic and nonphysical. Of course this is not new but the issue here is how avoid these nonphysical phenomena. As I can see these situations is very common in many physics areas, which surprised me strongly. In this particular case concerning the motion of light we have the $4-$four quantity vector $ \mathrm{d}k^\alpha=0$, this condition have to be modified when there is a gravitational field in the region where light it moves.\\[3 pt]
  
  Let's now clarify the meaning of the constraint condition related with the constancy of speed of light, which was used in the article \cite{martinez2022influence} to predict a possible variation of the wavelength of the light moving through a homogeneous and static gravitational field.
  \begin{equation}\label{equ23}
  \lambda_0 \omega_0=\lambda_1 \omega_1=\lambda_2 \omega_2= \cdots =\text{Constant}
  \end{equation}\\
  
 Now equation [\ref{equ18}] tell us something interesting. These equation can be used to obtain the shift in the frequency of the light, however there is no mention about whether or not the speed undergo a variation during its motion, as Einstein thought it could be possible and uses it in order to obtain the expected angle deviation as  can be read in his $1911$ article, on the other hand, the same result is possible using the fact that \texttt{actually the propagation of light is an interaction process with gravity} of course that a full description perhaps would be given using quantum electrodynamics (QED) but that idea is far from the scope of this article.\\[3 pt]
 
 Equation [\ref{equ23}] is basically a constrain imposed to the motion of light therefore along its path this must be satisfied, resulting that $c$ remains constant during its motion at every moment of time and points. Also can be applied to different observers due to the speed of light is a constant in special theory of relativity, is say:
 \begin{equation}\label{equ24}
\lambda \omega=\lambda^{\prime} \omega^{\prime}=\lambda^{{\prime}\prime}\omega^{{\prime}\prime}= \cdots=
 \end{equation}\\

Meanwhile equation [\ref{equ18}] can be applied only in two points, where is the source of light, the emitted, and in the arrived point. Nevertheless, this equation together with equation [\ref{equ24}] can describe \text{all the physical process of the propagation maintaining constant always the speed of light}. Of course not all physical process can describe using this two ideas due to that there is not use of the electromagnetism ideas, which is an covariant-theory concerning the Lorentz transformation. At this point I understand the issue about whether or not light it could undergo variation in its speed however is much natural and advantage suggestion that this should be remains constant if this is also true in the general theory in its full frame-work is something that we need to think of. Concerning this issue, I can say that also in the general theory of relativity equation [\ref{equ23}] could be modified or by some reason could be remains in that form. For me is amazingly that the electromagnetism is left out somehow by the theoretical frame-work of the general theory of relativity  which its the foundation and only the tensor energy-momentum is the unique link between them, $G_{\nu \mu}=\gamma T_{\nu \mu}$. As we have mentioned in the first article.\\[3 pt]

Following the idea of the unchanged of speed of light, from the computations made in the article \cite{martinez2022influence} we expect a variation in the wavelength apart of a frequency shift variation in the motion of light in a gravitational field which was not considered by Einstein's $1911$ paper. This variation as was calculated is: 
\begin{equation}\label{equ25}
\begin{aligned}
\dfrac{\Delta \lambda}{\lambda_2}=-\dfrac{\phi}{c^2}
\end{aligned}
\end{equation}\\[3 pt]

However, What does mean such expected variation of the wavelength? Is more common to find out a change in the frequency but equation [\ref{equ25}] could be used to calculate the speed of light which value we already known, $c$. A single source is not enough to do this so we need two source of light with the same wavelength and frequency and compare the displacement of the two wavelength at one point in the gravitational field. Doing the computation to this, we can get:
\begin{equation}\label{equ26}
\begin{aligned}
\dfrac{\lambda_2-\overline{\lambda}_1}{\overline{\lambda}_1} &= \dfrac{\phi_1}{c^2}\\ \text{And concerning frequency, we have:}\\
\dfrac{\overline{\nu}_1-\nu_2}{{\nu}_2}&=\dfrac{\phi_1}{c^2}.
\end{aligned}
\end{equation}

\begin{wrapfigure}{L}{0 pt}

\begin{tikzpicture}
\draw node[above= 3pt]{$\phi_1$}(0,0)--(2,0)--(2,-2)--(4,-2) node[above] {$\phi_0=0$};
\draw [dotted, line width=1 pt](3,-2)--+(0,4);
\draw[dotted, line width=1 pt] (1,0)--+(2,2)  node [ right=3 pt]{$\phi_2$};
\draw [dotted, line width=1 pt](2,0)--+(2,0);
\end{tikzpicture}
\caption{Two source light emission with the same $\lambda \,$ and $\nu$.}{\label{Fig:1}}
\end{wrapfigure}
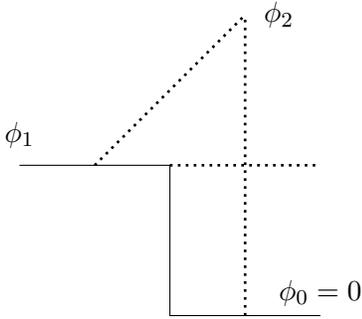

As can be see in figure (\ref{Fig:1}) we divide it  into two components with different values of the potential gravitational but with the condition that the $\nu$ and $\lambda$ of both sources is the same. Of course, we found that the result depends only on the $\phi_1$ potential. Therefore we can expect a little bit displacement on the wavelength at the point of potential in question. All this is undoubtedly very well-known but what is differ, is the use of the equation:
\begin{center}
$\omega=\dfrac{\gamma}{\left(1+\dfrac{\phi}{c^2}\right)}$
\end{center}

And the constrain:

\begin{center}
$\lambda \omega=\lambda_1{ \omega_1}= \cdots=\text{Constant}$
\end{center}

What I need to stress and clarify in more detail is the physical concepts contain  in equation [\ref{equ1}] as we used it here in this context, because I am afraid that could be misunderstanding or could be wrong or either both.\\[3 pt]

So in order to do that let me first of all, I need to clarify in a plain form what was the real purpose in propose this approach even when could be possible wrong or simply is a wrong manipulation of the idea what I had in mind. In other way the physical ideas which is based on could be wrong.\\[3 pt]

Even though equation [\ref{equ18}] is the limit case obtained from general theory of relativity, the derivation was made in a different way. Using the conservation energy law and  the propose that really the propagation of\texttt{ light is an interaction between gravity and using the corpuscular side of the electromagnetic field}, was able to arrive at the same equation, however in order to use it is necessary to impose the constraint, is say; the constancy of the speed of light as main requirement. \\[3 pt]

Concerning this, I would like to draw the attention to a particular article found in \cite[$283-295$]{kuhnelt1992teaching} which gave a demonstration of the Lorentz's transforms without using the propagation signal like light as Einstein did in his special theory of relativity. Of course, this is not a contradiction of Albert's theory and ideas. All the demonstration in\cite{kuhnelt1992teaching} is concerning that using only classical ideas is possible to derive the Lorentz's transformations but what I would like to stress here is the following: \texttt{If we request to Nature that all observers are equivalents} in the sense as Einstein taught us, the only sensible conclusion is that the speed of light have to be constant for all observers as well as the  inertial  ones. \\[3 pt]

Let me copy the transforms as general as possible I can, which I think contain some interesting physical aspects :
\begin{equation} \label{equ27}
\begin{aligned}
x&=\dfrac{x^\prime+u{t^\prime}}{\sqrt{1-Du^2}}\\
t&=\dfrac{Du{x^\prime+t^\prime}}{\sqrt{1-Du^2}}\\\text{Where $D$ is:}\\
\dfrac{\left(1-d^{-2}(u)\right)}{u^2}&=\dfrac{\left(1-d^{-2}(v)\right)}{v^2}=D=\text{Constant}\\
\end{aligned}
\end{equation}\\[3 pt]

Therefore if we take into consideration that, this transformation laws must be satisfied by the electromagnetic phenomena and the only possible and sensible choice is that \texttt{the speed of light have to be constant}, i.e,$D=\frac{1}{c^2}$. However, in spite of this I feel the necessity of a reason, a physical argument how this it so, is say; why in Nature exists a limit of information transmission from one point to another even if this is a local phenomenon and can not be applied to a global structure. Whether the constancy of the speed of information transmission remains constant a more large scale or not, is something that the current theories can not be give an answer to it as far I know and understand.\\[3 pt]

Which is surprisingly interesting is how A. Einstein in his $1911$ paper \cite{einstein1911influence} did the suggestion about the variation of the speed of light without gave it a sound  physical argument. At least all the above written point out that the speed of light is essentially a constant magnitude but I am not talking in relation of its numerical value but rather the fact that the constancy of the information transmission from one point in space-time to another so based on this, the following equation has not physical meaning:
\[c=c_0\left(1+\dfrac{\phi}{c^2} \right)\]\\

Moreover is possible that if there is a defensible unification at least in first order between the electromagnetic and gravitational fields, the constancy of information transmission should remains as a firm foundation to build up such frame-work for unify the two tangible fields that physicists know very well in some degree. To finish this part of the article is important to say that the motion of any body on the gravitational field is along a curve so called the geodesic, so light as the general theory say must be follow the same path which is determined by the source of gravity field. Of curse this is very understood by all the physics community. Therefore sooner or later in future paper will be necessary to talk about the energy-momentum tensor, which is symmetry, adding some useful term, and satisfy the condition: $\partial_\alpha T^{\alpha\nu}=0.$\\[3 pt]
\begin{align*}
T^\alpha_\nu=\fdv{\mathscr{L}}{{\partial_\mu}\phi}{\partial_\nu}\phi-{\delta^\alpha_\nu} \mathscr{L}
\end{align*}\\[3 pt]
Which is in general not a symmetry energy-momentum tensor but for the moment is not necessary to talk about it in great detail because this will take us beyond the scope of this little article.

\section*{\S $9.$ \, Equivalence Principle and the Motion of Light\\ The Quantum Aspect of the Electromagnetic Field}
Without doubt the equivalence principle is the foundation of the general theory of gravity with some limitations of course, however the relation with the constancy of the speed of light is hidden in some manner. When Albert Einstein made the thought experiment to show the possibility of bending  light, the equivalence principle play a secondary role since the effect of bending light is a consequence of its finite speed. On the other hand, the equivalence principle showed that this physical process should happen also in a gravitational field, that is what really is all  about the relation between the constancy of light and the principle mentioned. If had not finite value the speed of light none of those effects will occur in Nature so was necessary the idea to pass all this as possible tenable effects in a presence of gravity.\\[3 pt]

If this is a sound argument also for the general theory of relativity, we need to think more, however is plausible that will be reliable also. We also need to consider the quantum theory where the classical notions undergo a fuzzy understanding, philosophically speaking at this moment and consider the level of math up to now exists a unification theory could be possible. The particular relation established between  the non-inertial reference system with gravity was a good idea, however there is not inclusion of the electromagnetism phenomenon and that is a weak point. On the other hand, the equivalence principle was not intended to describe the electromagnetism, is an axiom to study gravity with all the ideas which came  from special theory of relativity.\\[3 pt]

So all the consequences that arises like the expected deviation angle of light, the shift frequency is basically a result to consider the speed of light or the information transmission as a fundamental constant independent of the inertial observers, the same argument is valid in a small region of a manifold where the special theory is remains valid. Of course, nothing of this is new but It is important to consider that a possible unification will need a revise of the main concepts in which is the founded the modern physics.\\[3 pt]

However, how about the quantum aspect of the electromagnetic field? I'm not talking about QED which a very sophisticated subject, what I'm referring is what A. Einstein pointed out in $1905$. Due to the rest-mass of the photon is zero, we can  use the relativistic equation $\mathcal{E}^2-p^2c^2=m_{0}^2c^4$ and find an equation between the energy and the momentum, so we have: 
\begin{equation}\label{equ28}
\mathcal{E}=pc
\end{equation}\\[3 pt]

From the point of view of the electromagnetism, there is a associated momentum with the electromagnetic wave and perhaps equation [\ref{equ28}] is an interesting relation between the special theory and the electromagnetism. Besides that from the $1905$ Einstein's paper we know that the energy of a single photon is in relationship with one of the characteristic of electromagnetic wave, the frequency $\nu$ in the follows way:
\begin{equation}\label{equ29}
\mathcal{E}=h\nu
\end{equation}\\[3 pt]

If we have a sound reason to equate both equation [\ref{equ28}] and [\ref{equ29}] then, we have:
\begin{equation}\label{equ30}
p=\dfrac{h\nu}{c}=\hbar \abs*{\vec{k}}
\end{equation}

So far, nothing new however this idea of light as a particle moving with a constant speed ,$c$, and the assumption of the interaction between gravity and the photon gave the same result got by A. Einstein in $1911$ paper. Therefore we can made the suggestion about the interaction between the photon and the gravitational field at point $x$ by means the following equation:
\begin{equation}\label{euq31}
\text{Basic interaction term} \rightarrow \dfrac{\hbar \omega}{c^2}\phi
\end{equation}\\[3 pt]

If we not taking into account other kind of possible interactions with the photon in the gravitational field, is normal to think that equation [\ref{euq31}] is the most simple and natural one to postulate. As consequence that $\omega$ is a function of $\phi$ we expect a deviation of the photon from its original path, that was the proposal made in the article \cite{martinez2022influence} in order to derive the equation:
\begin{equation*}
\omega=\dfrac{\gamma}{\left(1+\dfrac{\phi}{c^2}\right)}
\end{equation*}

Of course, there is not doubt that this isn't new however what is new is the manipulations made with this idea and the ability to uses it in order to obtain useful physical results as long as we accept the fact that conservation of energy for the case of the photon interacting with the gravitational field is valid as we used it in the last paper. Now, go beyond this proposal and try to use more sophisticated math and physics concepts is for now not possible at least for me. But is useful to think that perhaps some applications in Cosmology is feasible, how affects the expansion of the universe, a possible minimum energy value of the photon? In other words nowadays there are many defiance in physics.

\section*{\S$10.$\, Closing Remarks}

The moving of the photon in a gravitational field must be an interaction process, so It's necessary to demonstrate this assertion somehow. Why must be in this manner? Other considerations like the propagation as a wave in a gravitation field yields us to consider possibly the variation of the speed of light, where this speed must be \texttt{a fundamental constant in Nature and besides that makes more simple some physical formulations}. Even though the speed must be a fundamental constant in Nature, will be necessary try to find out a feasible justification about this constancy, I'm not talking about the number assigned to the speed but rather the reason why there exists a limitation in the information transmission from one point to another, perhaps a geometrical reason will be possible or could be found when the unification between the well-known physical fields happens. Try to point out another form to visualize the physical phenomena is very important more than ever because physics without new augment glasses could be a wrong way to follow.\\[ 3 pt]

A possible minimum energy value of the photon could gives us perhaps a new insight concerning the expansion of the universe. However, take this statement seriously needs more physical justification rather than a mathematical frame-work because nowadays most theoretical ideas does not have a good physical reason to consider it as nice and elegant way to follows. Photon can not have either higher value as much as we want or a lower value than a minimum one, what is left pending now is to figure  out that and apply it to cosmology for instance. As we know the task is not easy and as consequence I'll try to study more deeply this issue but always taking into consideration the basis ideas where the frame-work were constructed. At least in my mind there is the idea,that really the motion of the photon in a gravitational field is an interaction process so to go beyond this and apply more physics concepts I need to clarify this mechanism and  consider quantum ideas. So in some point of this research I need to see how this process could be derived from QED. So, the process of \texttt{ interaction between gravitational field and others fields is the idea and keeping always constant the speed of light} could be the two guides in order to explore what we can get from this but now in higher use of physical concepts.\\

\section*{\S$11.$\, Corrections to Some Equations of the Article \cite{martinez2022influence}}

In the previous article  \cite{martinez2022influence}, we had been established that the energy of a photon in a gravitational field as follows:
\begin{equation}\label{equ32}
\begin{aligned}
\mathcal{E}=\hbar \omega+\dfrac{\hbar \omega}{c^2}\phi
\end{aligned}
\end{equation}

From  equation [\ref{equ32}] we obtained the relation between the frequency at two point where exists a gravitational field and deduced from it the wavelength displacement as well as the frequency shift without using the suggestion of the variation of the speed of light. Whether this is a useful concept or not I can not tell anything concerning this issue but turns out to be more easy to keep constant the speed of light as a feasible argument. Also we take advantage of the idea of interaction to write a energy conservation law along the path followed for the photon, is say:\\[3 pt]
\begin{equation*}
\begin{aligned}
\mathcal{E}_1=\hbar \omega_1+\dfrac{\hbar \omega_1}{c^2}\phi_1=\hbar \omega_2+\dfrac{\hbar \omega_2}{c^2}\phi_2=\hbar \omega_3+\dfrac{\hbar \omega_3}{c^2}\phi_3= \cdots= \cdots \text{Constant}
\end{aligned}
\end{equation*}

The equation [$9$] in the previous article there was a error in the subscript, so it must have been as follows:
\begin{equation*}
\begin{aligned}
\dfrac{\lambda_1-\lambda_2}{\lambda_2}=\dfrac{\phi_1-\phi_2}{c^2}=\dfrac{\phi_2-\phi_1}{c^2}, \, \abs*{\phi}>0
\end{aligned}
\end{equation*}
\begin{center}
Submitted date: (September $16$, $2022$)\\
Second Article
\end{center}
\begin{center}
\begin{tikzpicture}
\draw (2,0) -- (11,0);
\end{tikzpicture}
\end{center}

\section*{\S $12.$ The Line Element for Light in a Weak Gravitational Field\\ A Little Introduction}
As we presented in the last paper \cite{martinez2022influence} concerning a certain comments about the A. Einstein's $1911$ paper \cite{einstein1911influence}, it has be proved there that a energy consideration and the guide idea of interaction process between the motion of the photon in a weak gravitational field led us to the same conclusion that Albert got and published in $1911$. In the article it was also argued that the suggestion of the changed of the speed of light was not necessary and it was established an ideal energy equation which led us to the relationship between the angular frequency and the potential of the gravitational field, is say: 
\begin{equation}\label{equ33}
\begin{aligned}
\omega=\dfrac{\gamma}{\left(1+\dfrac{\phi}{c^2}\right)}\\
\abs*{\phi}>0, \, \phi \rightarrow \, 0 \, \text{at} \,r  \rightarrow \,\infty \\ \text{Using the following equatio:}\\
\mathcal{E}=\hbar \omega+\dfrac{\hbar \omega}{c^2}\phi\\
\end{aligned}
\end{equation}\\[3 pt]

What was left out in the paper is about the line element of the photon, It is know that this must be zero, $\mathrm{d}s=0$ so if the idea presented has a physical meaning at least this result could be derived from it. I comprehend that this must be uncommon in some sense but if the idea of a real interaction is true it is necessary to study all its consequences, at least in the classical physics and is a good idea to do this however  more profound study of the idea presented in the paper have to be a must.

\section*{\S $13.$ Energy and Line Element of the Photon}

If we ask ourselves about the motion of the photon, we speak a single photon but we know that  is for the sake of simplification, it is possible to established a relation between the energy due to the changed in the angular frequency and the changed in the gravitational potential along the path followed for the photon through its motion, is say:\\[3 pt]
\begin{equation}\label{equ34}
\begin{aligned}
\hbar \mathrm{d}\omega=-\dfrac{\hbar \omega}{c^2}\mathrm{d}\phi
\end{aligned}
\end{equation}\\[3 pt]

Taking into consideration that the photon is moving along a curve line in the space, we can divide  the right-hand side of the equation [\ref{equ34}] by the arc element, $\mathrm{d}\overline{s}$, covered in the time $t$ where the arc element  is not the four-dimensional arc element,$\mathrm{d}s^2=g_{\nu \mu}\mathrm{d}x^{\nu} \mathrm{d}x^\mu$, defined in the general theory of relativity,  in other words:

\begin{equation}\label{equ35}
\begin{aligned}
\mathrm{d}\omega=-\dfrac{\omega}{c^2} \dfrac{\mathrm{d}\phi}{\mathrm{d}\overline{s}} c\mathrm{d}t\\ \text{Where}\\
\mathrm{d}\overline{s}^2=\mathrm{d}x_1^2+\mathrm{d}x_2^2
\end{aligned}
\end{equation}\\[3 pt]
Considering the equations [\ref{equ35}] and [\ref{equ33}], we can firstly differentiate the first equation of [\ref{equ33}] to obtain a relation between $\mathrm{d}\phi$ and $\mathrm{d}\omega$ and substituting it in the equation [\ref{equ35}], so:
\begin{equation}\label{equ36}
\begin{aligned}
\mathrm{d}\omega=-\dfrac{\gamma}{c^2}\dfrac{\mathrm{d}\phi}{\left(1+\dfrac{\phi}{c^2}\right)^2}, \, \text{and}\,\, \omega=\dfrac{\gamma}{\left(1+\dfrac{\phi}{c^2}\right)}\\
\mathrm{d}\overline{s}=\left(1+\dfrac{\phi}{c^2}\right) c\mathrm{d}t
\end{aligned}
\end{equation}

Paying attention that:
\[ \mathrm{d}\overline{s}^2=\mathrm{d}x_1^2+ \mathrm{d}x_2^2\]

Taking the square of both sides of equation [\ref{equ36}] and compare them, we have then: 
\begin{equation} \label{equ37}
\begin{aligned}
\mathrm{d}x_1^2+\mathrm{d}x_2^2=\left(1+\dfrac{\phi}{c^2}\right)^2  c^2\mathrm{d}t^2 \approx \left[1+2\dfrac{\phi}{c^2}+\order{c^4} \right]c^2 \mathrm{d}t^2
\end{aligned}
\end{equation}
 
 So, we can rearrange the equation [\ref{equ37}] in order to encounter the very well-known line element of the path followed by the light in the space-time framework nearby in a weak gravitational field. This result found here by itself is very interesting and in my consideration have to be study whether is true or not, however this is only my humble opinion and this is what I could point out, to be demised by researchers in this field with more knowledge in experience. So, we have definitely:
 \begin{equation*}
\begin{aligned}
 \mathrm{d}x_1^2+\mathrm{d}x_2^2&= \left(1+2\dfrac{\phi}{c^2} \right)c^2 \mathrm{d}t^2\\ \text{Or in more familiar way:}\\
 \left(1+2\dfrac{\phi}{c^2} \right)c^2 \mathrm{d}t^2-\mathrm{d}x_1^2-\mathrm{d}x_2^2 &=0\\ \text{Which is basically: }\\
 \mathrm{d}s^2&=0
\end{aligned} 
 \end{equation*}
 \vspace{4 cm}
 \begin{center}
 \vspace{-3 cm}
 \end{center}
 \begin{center}
\begin{tikzpicture}
\draw (4,0) -- (11,0);
\end{tikzpicture}
\end{center}
\begin{center}
Submitted date: (September $24$,\,$2022$)\\
 Third Article \footnote{I'd like to address the issue that may be the derivation  has not the force of a prove in physics but is a first insight about how could be obtain the line element for the light in space-time, of course that this not contradict anything done in the previous article, version 3}
\end{center}
\pagebreak
\section*{\S $14.$ \, Ideas Concerning On the Variation of the Speed of Light\\ Albert Einstein, M. Abraham and H. Weyl\\ $1911,\, 1912 \, \text{and} \, 1918$\\\center{Second and Last Part}} 
 The main idea in this opportunity is to talk about the following:\\
  
  As we know the speed of light is a physical fundamental constant in Nature, as we mentioned many time in \cite{martinez2022influence}, but this was not always the case, particularly in the seek for a new theory of gravity; two physicists made the conjecture about a possible variation in the speed of light in order to obtain suitable equations with physical meaning but it was not fruitful. With this in mind, one can imagine most of the simple variations using Wely's idea concerning the generalization of Riemann's Geometry. In this section, we are going to study these ideas and make a few comments in relation to the utility of the idea in those years and its uses as a model to obtain field equations for the gravity problem raised by A. Einstein.

\subsection*{\S$14.1$ Introduction to the puzzle, why the persistent idea of the variation of the speed of light?}

\epigraph{ \textit{$\cdots$ both phenomena, gravitation and electricity, have remained completely isolated from one another up to now\\ \textit{H. Weyl}}}

   In the article \cite{martinez2022influence}, version $6$, we  made two basic assumptions and from them we were able to demonstrate that was no necessary the proposal made by Albert, concerning the variation of the speed of light in his $1911$ article \cite{einstein1911influence}. The idea of energy conservation law, the constancy of the speed of light and the crucial idea what really happens is an interaction process between the gravitational field and the corpuscular side of the EM field, this permits us to obtain all the physical phenomena got by A. Einstein in his article. This tells us that the assumption of the possible variation of $C$ was no a good idea at that time, however in the next publications until $1916$ Einstein strongly insisted in establish a gravitational theory on this unnecessary idea so we ask ourselves, Why?\\
   
  In the article published in $1912$ ,titled: \textit{The speed of light and the statics of the gravitational field} \cite{einstein1912speed}, he found a way to get a field equations based on his ideas but considering the variation of the speed of light as the unique foundation which seen to be reasonable to him, however was an incorrect approach as he proved later on. The basic field equations found has not physical meaning at all, but we are going to begin with that:
  \begin{equation} \label{equ38}
\begin{aligned}  
  \laplacian\Phi=kc^2\rho
  \end{aligned}
  \end{equation}
  
  What was the meaning of equation [\ref{equ38}]? In order to give a physical answer to this question we need to recall what we have done in the first article \cite{martinez2022influence}, where the two of the most interesting physical assumptions were to consider the process as an interaction between gravity and the quantum aspect of the EM field, and second; the idea of keeping constant the speed of light in a presence of gravity.Then, we used the energy conservation law to derive the equation that relate the frequency $\nu$(angular frequency $\omega$) and the gravitational potential. If we take the idea seriously,the constant of the speed of light, and we stay true about that we can easily say that equation [\ref{equ38}] is incorrect and has not physical meaning in any sense. However, this argument is not sound and we must give more reasonable arguments to consider equation [\ref{equ38}] as non-sense. Beforehand is a good idea to mention that is will not be an easy task to carry out.\\
  
  To Albert in the years $1911-1912$, the only possible solution to the puzzle was to consider the speed of light as a scalar field $C=C(x)$, which has to have the main role and much more than that it controls all gravity phenomena as we can see in equation [\ref{equ38}]. The idea was no new of course, but the approach was the main thing in $1912$ to Albert. Therefore, the conclusion was that the variation of the speed of light was really the true nature in relation to the gravitational phenomenon which we already know that is not the real way to think about it. All this took place due to Albert in those troubles years, he didn't take into account the geometry interpretation of the Lorentz's transformations found by Herman Minkowski in $1909$. So was only taking into account one component of the metric, speaking in the modern language even when Albert didn't know that, the temporal one but in the wrong way, is say: $C$ was a dependence function of the gravitational potential $\Phi$. Let's write down for future reference;\\
  \begin{equation} \label{equ39}
\begin{aligned}  
  \mathrm{d}s^2=g_{\mu \nu}\mathrm{d}x^{\mu}\mathrm{d} x^{\nu} \rightarrow g_{00}\mathrm{d}x^0-\mathrm{d}x^1-\mathrm{d}x^2-\mathrm{d}x^3=\mathrm{d}s^2
  \end{aligned}
  \end{equation}
  
The reason why Albert didn't take into consideration the temporal component correctly of the metric tensor [\ref{equ39}] was because he didn't pay much attention to the geometrical interpretation given by H. Minkowsky in $1909$, and in his mind still persisted to model the new theory of gravity using the Newtonian celestial mechanics,see equation [\ref{equ40}], which from his point of view of the problem was a suitable solution, is say: he expected that the new theory has to have a similar mathematical structure but there were two  physical quantities which would determine the gravitation potential as we can see in equation [\ref{equ38}], which now is known that it was far away from the true nature of gravity. Whereas all mentioned above was different from Newtonian's idea of gravity and therefore was not congruent with any ideas from the special theory of relativity. 
\begin{equation}\label{equ40}
\laplacian \phi=4\pi G \rho
\end{equation}

Of course, as we know now this was not the correct physical assumption in order to obtain useful field equations for the gravitational field. From these points we are going to make a more deeply analysis to the Albert's paper and focus us on the reason why the velocity of light, its variation was so important in order to build up the new theory, was essential. It is good to remember that all the results obtained by Albert in $1911$ could be derived using energy conservation law as we demonstrated in the previous article \cite{martinez2022influence} and some other assumptions. Also we need to consider that in the $1911$ paper Albert made a crucial conjecture concerning the variation of the speed of light without given any physical arguments to deduce an angle deviation in first order approximation, in which is not necessary the equivalence principle, and then in $1912$ wrote a paper where assigned to the variation of the speed of light as the basis to underlie a possible theory of gravity. All that suggests in my opinion that it was a logical error because there was no a physical reason to postulate that, the matter is why he did it anyway.\\

Besides, pay attention that in $1911$ publication, all the rest physical deductions were made without consider the variation of $C$, which to me is no physical and not logical because he just made used of the variation of $C$ where he needed it. No consider equation [\ref{equ39}] as the real basis for a plausible field theory, however what is out of the place in my opinion is to take seriously that $C=C(x),\, x=x(t)$ as a good candidate for the field theory of gravity and for that reason we're going to study carefully the Einstein's idea in order to look forward from his perspective. Before doing that, I see necessary to study a little more both Einstein and Max Abraham's \cite{abraham2007new} ideas about the variation of the speed of light. Notice that Einstein already has made mentioned of Abraham's paper in his own publication in $1912$ so he knew the work made by Abraham around the possible variation of the speed of light as a good way to solve the gravity problem. Which in physical terms has no real solution, of course we can not justify Abraham's point concerning his new theory of gravity in $1912$.\\

\textit{First ideas to the wrong path$ \cdots$\\}

  \textbf{From the constancy of the speed of light to the conjecture of its variation due to Einstein puzzle with Gravity, What was the cause of that?}\\
  
  Since $1905$ and due to all physical experiments made in order to measure the speed of light with respect to the hypothesis of the ether, pointed out to the conclusion that the speed of light is a true constant in Nature like the gravitational constant in the Newtonian theory of gravity, $G$. Therefore, Albert was able to construct a theory which has changed all the Newtonian concepts of space and time as absolute physical manipulations into a more dynamical quantities. Moreover, the constancy of the speed of light was the building block to made that. \texttt{The speed of light in vacuum is the same in all possible directions and for all possible inertial observers}.\\
  
  From that beautiful statement was deduced all physical consequences known now as the special theory of relativity. Even when there was no reason why such a speed must has to be constant, that statement was taken as the foundation of the modern physics, so physicists from that moment up to now only just saying: $c$ \texttt{it is a physical constant magnitude in Nature} \footnote{I have the idea that shortly will be found a geometrical reason for that constancy}. For that point, is strange to make a conjecture about the variation of that constancy in order to find out a mathematical model for the gravitational problem raised by the special theory of relativity. However, despite all that what's the meaning of that idea? and much more the meaning of the following equation, physically speaking:
  \begin{equation}\label{equ41}
\begin{aligned}  
  c=c_0 \left(1+ \dfrac{\Phi-\Phi_0}{c^2} \right)
  \end{aligned}
\end{equation}  

First of all, equation [\ref{equ41}] is just a first order approximation, we can read that in both articles by Albert and Abraham, $1911 \,\text{and}\, 1912$. Therefore, both assumed immediately that light in a presence of a gravitational field propagates as wave phenomenon in turn was possible to obtain an angle deviation of light using equation [\ref{equ41}], that was the whole idea concerning to Einstein and his equivalence principle. This turned out to be a successful idea, but really was a wrong approach to the solution of the problem. As far as I can see, equation [\ref{equ41}] has not any physical meaning at all, I have said, and besides the dependence on $C$ concerning $\Phi$ turns out to be unnatural and forced to get what was got in $1911$ by Einstein about the angle deviation and justified later on in his $1912$ paper only. However, in his book published \cite{einstein1948special} in $1916$ the idea of the variation of the speed of light was abandoned partially.\\

One of the objections in favor of the variation in the speed of light was the following: If a point-like particle moves in a gravitational field will gain or lose potential energy, so if we assume Einstein's famous equation could be applied to this situation we expect a variation of one of the two terms in the equation, indeed Abraham himself mentioned this in his article \cite{abraham2007new}.
\begin{equation} \label{equ42}
\begin{aligned}
\mathcal{E}=\gamma m_0 c^2,\, m=\gamma m_0
\end{aligned}
\end{equation}

Then, the idea is that $C$ has to undergo a variation in its value in the presence of a gravitational field which in turn it is produced by itself. The argument stated physically speaking has no meaning, as I mentioned many times above, but the fact mentioned would be of interest in the sense that if equation [\ref{equ42}] is valid for all kinds of energy, including the potential energy there will be something to think of. In addition of what was mentioned, we could read in \cite[page~181]{matveev1976mechanics} the following:
\begin{quotation}
 The energy conservation law as expressed by [\ref{equ42}] indicates that this relation is quite likely to be valid for the potential energy, i.e. that the potential energy has inertial property.
\end{quotation}

This suggests to us that really when a point-like particle, a photon for instance in our study case, is moving in a gravitational field is most probably that be the mass which undergo a variation due to a gain or lose in the  potential energy and in addition to that it is pointing out something that we proposed in the previous article \cite{martinez2022influence}, that what's really happens when a light-ray pass nearby a heavenly body is an interaction process which is contrary what Albert and Abraham suggested in $1912$. In the case of the photon must undergo a change in the frequency which equation could be found using the energy conservation law as we did in \cite{martinez2022influence} using the fact that we are dealing with a conservative field. So, is more logical to keep constant the speed of light and try to obtain a new theory of gravity using other ideas, and that was done by Albert himself when he finished his work in $1916$ and by Hilbert also.\\

Hence, we gave a physical and sound argument in order to invalid the idea of M. Abraham concerning his believe of the variation of the speed of light expressed by equation [\ref{equ41}] as a possible foundation for a new theory of gravity, moreover it is interesting to pay attention to the quotation given because is giving a plausible inertial property to the gravity through equation [\ref{equ42}] if could be valid for all kinds of energies, so this was basically what we had in mind when we stated that all boiling down to an interaction process. Also it is interesting to mention, that equation [\ref{equ42}] is a first order approximation obtained from:
\begin{equation}\label{equ43}
\begin{aligned}
\dfrac{c^2}{2}-\dfrac{c_0^2}{2}=\Phi-\Phi_0\\
c_0^2=c^2-\left(\Phi-\Phi_0\right)\\
c_0=c \sqrt{1-2\left(\dfrac{\Phi-\Phi_0}{c^2}\right)}\\
\end{aligned}
\end{equation}

\textit{ Second idea to the wrong path $\cdots$, we expect something different$\cdots$}\\

Now, Abraham made a mathematical trick to get the same relation as Einstein got in his publication in $1911$. It is a kind of mathematical trick because the dependence could be obtained directly from the basic first equation of [\ref{equ43}] but he didn't. Though is possible it is necessary the constraint concerning the ratio $\frac{\Phi}{c_0^2} \rightarrow 0$ where $c_0$ is the value of the speed of light at the origin of the coordinate system, based on the original Abraham's idea. In the third equation of [\ref{equ43}] M. Abraham like Einstein, also made use of the constraint $\frac{\Phi}{c^2} \rightarrow 0$, however as $\Phi$ is supposed to be a function of $C$ has to have a first order dependence $\Phi=\Phi(C)$ and that is something to take into account, turns out to has physical sense but really is not so if we spent a little bit more time thinking in the subject mentioned above.
\begin{equation}\label{equ44}
\begin{aligned}
c=\dfrac{c_0}{ \sqrt{1-2\left(\dfrac{\Phi-\Phi_0}{c^2}\right)}}\\
c \approx c_0 \left[1+\left(\dfrac{\Phi-\Phi_0}{c^2}\right)\right]
\end{aligned}
\end{equation}

If from the first equation of [\ref{equ43}] we isolated the variable $C$, we obtain a complete valid mathematical relationship, is say:

\begin{equation}\label{equ45}
\begin{aligned}
c=c_0 \sqrt{1+2\left(\dfrac{\Phi-\Phi_0}{c_0^2}\right)}\\
c \approx c_0\left[1+\left(\dfrac{\Phi-\Phi_0}{c_0^2}\right)\right]
\end{aligned}
\end{equation}

Equation [\ref{equ45}] could be valid if the constraint at the origin is also true and that would be a contradiction to Abraham's idea and in turn to the proposal of  the variation of light made in $1911$ by Einstein. Therefore, would be interesting to give some physical insight if the statement could be taken seriously. The main point is, what was the real idea to say that the variation could determine the gravitational field and to deduce from that ideal field equations. In Einstein's paper \cite{einstein1911influence}, the only use of the variation of the speed of light was to compute the angle deviation and all the rest of the physical deductions were made without this assumption so the matter here is, why he considered that as a tenable tool to get what he thought first using the equivalence principle which is not necessary for that.\\

We have to remember that the deviation of a light-ray is due to the finite speed, its inner property of light, so the use of the equivalence principle is to correlate both phenomena, is say, an accelerating reference systems are equivalents to an inertial observers in a constant gravitational field. Therefore, is more natural the approach made in \cite{martinez2022influence}, using at the same time the two postulates made concerning keeping constant the speed of light and the interaction process that might occur when a ray of light pass nearby a heavenly body like the sun for instance. Now, to go from this to formulate,perhaps, a more complete gravitational equations is far away from my own possibilities but I must try out and that is the real thing concerning to write papers touching physical ideas which now is very common among the specialists in the subject.\\

Keeping in mind all the above mentioned, we are going to study even further both articles written by A. Einstein and M. Abraham in order to give more physical and sound arguments in favor of keeping constant the speed of light even in the presence of a gravitational field including also in relation with the general theory of gravity. Therefore, we could perhaps say that there's something that must to be added to the idea of gravity as a geometrical theory. Following the last idea, the problem as always is how we can do that knowing that there are some many misunderstanding concepts like space and time and that raise many questions. For instance, the real meaning of space and time when one think about the  unification problem and some ideas from the quantum gravity approach, for more information, please see: \cite{rovelli2006unfinished} where the reader will encounter some insights concerning this difficult task in theoretical physics.\\

\textit{We are lost in the center of the universe $\cdots$}

\epigraph{
\textit{The speed of light, $C$\\ Take its variation and made a new theory of gravity}}{A. Einstein\\M. Abraham\\ \texttt{Think up a new theory of gravity from nothing}}

Let's now begin with the most important discussion concerning this point made by these two revolutionary minds. Also, we're going to roughly study what H. Poincare and Lorentz thought about this issue. Let's start with Albert Einstein, of course, and his attempts to successfully get a theory of gravity compatible with the special theory requirements.\\

The year is $1905$, A. Einstein have already realized that the speed of light is the same in all possible directions in $\mathbb{R}^3$ and for all possible inertial observers which agree with the theory of Electromagnetism. In $1909$ H. Minkowsky published a paper\cite[page,$75-91$]{lorentz1952principle} where gave a geometrical interpretation of the Lorentz's transformations but A. Einstein did not pay much attention, for further historical information see \cite{logunov2004henri}, finally the beginning of the seek for a genuine theory of gravity began in $1911-1916$. When Albert tried to face the problem with the idea of a scalar option being suitable for him and assuming that this scalar field had a relationship with the possible variation of the speed of light, which turns out to be the only physical quantity to be promoted a scalar function by Einstein and others, he did not notice the major physical problem involved working with that kind of idea.\\

In $1911$ he used the idea that light propagates as a wave and the function dependence $C=C_0+a \times x$, it permitted him to deduce an angle deviation in first order but he didn't give a physical meaning, why have to be in that way, in fact he didn't use the idea to obtain the other effects like the shifted of the frequency in the light which was emitted by a light source. Energy conservation law and the statement of an interaction process between the corpuscular side of the EM field and the gravitational field yields us to the same physical results as we demonstrated in \cite{martinez2022influence} \footnote{Please, it is a good idea to study carefully this article apart from some possible grammatically error pay a lot attention to all the physics concepts used which are completely different and attractive physically speaking.}. The scalar interaction term could be written as: \[h\nu\dfrac{\phi}{c^2}\]\\

 Here we \texttt{assumed that is the mass which undergo a variation} and not the speed of light, which is contrary what M. Abraham thought in $1912$ like Einstein as well. In other words, photon doesn't feel any curvature around the heavenly body when pass nearby one. That's might be the cause why the Electromagnetism equations written in $4-\texttt{D}$ is invariant when we use the covariant derivative, is say \footnote{{In what sense really this statement has a physical meaning is the key idea to comprehend the reason why I was able to write the article in $2022$, something is missing in Einstein's idea concerning the gravity phenomenon}} the main thing is looking for it and be capable to add it into the Einstein's field equations but that was not possible in any sense. Let's see the Maxwell's $4-$dimensional equations in which the main role is play by the $4-$potential, $A_\mu$.
\begin{equation} \label{equ46}
\begin{aligned}
F_{\nu \mu}=\partial \Phi_;\nu-\partial \Phi_;\mu=\partial\Phi_,\nu-\partial\Phi_,\mu
\end{aligned}
\end{equation}\\

Equation [\ref{equ46}] contains as a fact the constancy of the speed of light encoded in it. So, it is not strange that it will be an invariant field equations because there is no a clear meaning how really gravity could affect the EM field and the other way around but the main idea is there, $C$ is an important  physical constant and we need to seek an approachable geometrical interpretation. The main point, in my opinion, consists in to find out a link, a physical link, between gravity and the electromagnetic field and that seems to be a huge barrier to jump.

\subsection*{\S $14.2$ \, A. Einstein's $1912$ paper}

Before we point out a few points in favor to the constancy of the speed of light instead of its variation, let's do the following thought experiment. Let $\phi$ be a scalar field, which nature we don't know yet, and we consider the following equations to be valid in the case when there is a gravitational field in the sense of the first idea presented by A. Einstein:
\begin{center}
\begin{equation}\label{equ47}
\begin{aligned}
u\phi=c_0, \\\texttt{being u the speed of light in the presence of the scalar field}\\
\mathrm{d}u\phi+u\mathrm{d}\phi=0\\
\mathrm{d}u=-u\dfrac{\mathrm{d}\phi}{\phi}\\
\mathrm{d}u=-u\psi_{m}\mathrm{d}x^{m}, \, \phi=\phi(x^m)
\end{aligned}
\end{equation}
\end{center}

Therefore, in the fourth equation of [\ref{equ47}], is possible to make a link with the Weyl's idea\cite{weyl1952gravitation} of the generalization of the Riemann's geometry. In other words, this is related with the variation of the length of a vector when is displaced parallel to itself in a differential manifold. Of course, we consider as always that in vacuum the speed of light is a constant physical magnitude, $C$. This is just an idea, but the point is not that but rather than to study closer all the physical statements in the A. Einstein's $1912$ article. Besides, equation [\ref{equ47}], is the most simple formulation to a possible variation of the speed of light if that was a real possible solution for a physical problem, keeping in mind those ideas and let's continue with the argument presented in $1912$.\\

When A. Einstein made the computation of the angle deviation he assumed that both, the propagation far away of any source of gravity and nearby is a wave phenomena and that permitted him to obtain what he was looking for. Thereby, the idea of a wave propagation was an ideal method to complete that task. however no physical meaning was given to support that conjecture, and because of  that I consider that the approach have not any physical meaning at all. In $1905$ was already known by Einstein the notion of a concentrated package of energy in a region, photon, but he didn't use that and preferred to formulate everything in thought experiments and was not careful concerning the significance of the constancy of the speed of light and its possible interaction with the gravitational field as we treated in the article \cite{martinez2022influence}. Therefore, we can say that from two postulates only, one was able to get all the physical phenomena keeping the logic concerning the main tangible property of light,\texttt{ its constant speed}. Which sounds more like a real physical progress in conceptual terms than the one proposed by Albert in $1911, 1912$. concerning what I think about this, is more sensible the idea of keeping constant the speed of light rather than to consider a possible variation without any sound physical argument or necessity.\\

But, in $1912$ Einstein wrote a paper with potential field equations for gravity. Did it make sense? If it was so, Why did not work the idea in the general theory of relativity as well? In fact, there is no such a thing in the foundation of the theory presented in $1916$ which the main role was given to the metric, for more further information see \cite{mcmahon2006relativity}, is say;
\begin{equation}\label{equ48}
\begin{aligned}
\mathrm{d}s^2=g_{\mu \nu}\mathrm{d}x^{\mu}\mathrm{d}x^{\nu}\\
\mathrm{d}s=\cdots\text{contant}\cdot \mathrm{d}\tau
\end{aligned}
\end{equation}\\
Where the constant is the speed of light,$C$, which is a fundamental physical quantity in the modern physics. But, Einstein took another path in order to get the field equations,namely, something with more physical intuition however without a sound reason for the statement made in his $1912$ paper. The main reason why this idea was abandoned later on was due to that the output of the computation was not the correct angle deviation as calculated using the general theory of relativity. Other reasons apart from what was mentioned is due to the physical principals of the special theory of relativity,namely, all the necessary requirements were no included in the equation [\ref{equ38}]. So $C$ underwent a variation without a sound reason but rather to obtain a scalar equation to make calculations, which is very common in new ideas to make a lot of error until it is possible to arrive at the right path.\\

It is understandable to imagine how difficult was to establish physical ideas in order to get the equations for the gravitational field, but make the assertion concerning a possible variation in the speed of light was more difficult and try to justify all the physics known at that time is an overwhelming task. In physics is very common that when one says something about the nature of light or the space-time one encounter the hardest mental barrier, \texttt{The common sense!}.\\

  \textit{From the center of the universe to a non-location in space or time $\cdots$}\\

Even when finally published his theory of gravity, he persisted in the possible variation of the speed of light, however in this opportunity he gave a good physical reason, mechanical one, let's read what he wrote in\cite[page,~68 ]{einstein1948special}, in the Spanish version:\\
\vspace{-0.6 cm}
\begin{quotation}
$\cdots$Secondly,the previous consequence proves that as the general theory of relativity indicates, that the such mention constancy law of the speed of light in vacuum-that is one of the two postulates of the special theory of relativity- can not be valid anymore, as consequence that the only way a light-ray could undergo a deviation from its rectilinear motion is a consequence of a change in its velocity which depends on the position\footnote{This is a translation made by the author of this article, from the Spanish language version of the book mentioned}
\end{quotation}

That is a good point, and is also true in classical mechanics but the idea is still there because the velocity as a vector can undergo a change but in the case of the light-ray its speed it is keeping constant which is the main idea. So, we must remember that velocity and speed are two different concepts in physics. Therefore, we need to give a sound physical reason in support the idea of the constancy of the speed of light even in the presence of a gravitational field. Why must be true? But, we need to focus ourselves on what Einstein wrote, the physical ideas, in his $1912$ article. Before we begin with the physical analysis, I would like to point out something very interesting concerning the angle deviation of a light-ray said by Einstein himself in his $1916$ books \cite{einstein1948special} where he mentioned that part of the angle deviation is due to the gravitational attraction and the other half of the value due to the curvature produces by the heavenly body where the light-ray pass nearby,\textbf{ therefore a question arises on this, What does that mean?}.\\

It is healthy to keep in mind this, in the sense that partially is giving me the reason concerning the interaction process between both fields. However, the problem is that there is only a scalar term, namely, there is no directly relation with all the components of the metric tensor, $g_{\nu \mu}$. whereby, all this thinking about gravity and the EM filed is pretty complicated but a solvable problem. Let's now post a question, What's the meaning of Einstein's idea concerning the change of the speed of light? The answer, none. At least, at this point where there is no unification between both gravity and the electromagnetic field. But classically is a good physical reason to postulate that even in the presence of a gravitational field, the speed of light must be a constant physical quantity, the problem with this is justify it. Moreover, all this leads us to consider other phenomena that could occur during the interaction which in turn gives a reasonable answer about the angle deviation experienced by the light-ray and perhaps other particles due to the double behavior,i.e, wave and particle behaving.\\

Le's now analyze the proposal made by Einstein in his $1912$ article.\\

In the article  $1912$ \cite{einstein1912speed} appears for the first time one of the approach to a non-inertial reference system but with the wrong point in Albert's mind, i.e. he tried out to investigate  for second order terms in the coordinates of space and time like a Tylor expansion, one can see this in the following equations: 
\begin{equation*}
  \begin{aligned}
\zeta & = \lambda + \alpha t^2 + \cdots\\
\tau & = \beta +\gamma t+ \delta t^2 + \cdots\\
\end{aligned}
\end{equation*}

In a small values around a point in space-time this expansion are completely valid, as Einstein's mentioned in his article, in the sense that he wanted get a classical mathematical relationship such as in the free fall body; he needed an acceleration in $\zeta$ and a first linear approximation in $\tau$ and it was so. The difficulties one can see with this expansion is that the constrain relation as Minkowski said, using the modern language, the differential quadratic form must be invariant and which is the mathematical expression for the constancy of the speed of light and in the Einstein's idea this is not the case in any sensible sense. Why? As Einstein pointed out, $C$ is a dependence function of the gravitational potential, which is constant with respect to time, so in this context the differential quadratic form could not be apply, $C$ is no longer a constant physical magnitude any more  in the sense proposed by Albert.\\

The misleading way perhaps could be due to that $C^2$ has the same unit as the gravitational potential $\phi$ so is an easy way to make a conjecture concerning a plausible variation and a link in the sense that both authors Albert and Abraham thought as approachable one solution, $[\frac{\phi}{c^2} ]\Leftrightarrow [1]$ from this was possible to obtain the equations with the requirements imposed by Albert, namely; in the end of this one can see that a relationship between the speed of light and the gravitational potential is obtained but in my opinion with no physical sense as I mentioned many times in this article. The rest of the article is just merely a justification of the conjecture but at  a higher cost, for instance, to consider that the gravitational constant is not more constant\footnote{see the article where is placed in discussion the possibility of the variation of the gravitational constant, which seems to me a wrong death way to solve any physical problems, \textbf{Cosmic String in the Tensor-Scalar Theories on Gravity, Noe Morales, UNM(Metropolitana, Iztapalapa Unit)}} and instead there is another one; so one can see that this theory is a good approach to the use of the metric tensor but at that time Albert did not notice the utility of this mathematical object.
\begin{equation*}
\left \{
\begin{aligned}
 \zeta & = x+\dfrac{ac}{2}t^2\\
\eta &=y\\
\zeta &=z\\
\tau &= ct \\
\text{One can obtain from  the above relations:}\\
c & = c_0 +ax
\end{aligned} \right.
\end{equation*}

One can see a familiar mathematical equation, so the idea have two mainly problems; first, use the differential quadratic form, mathematical representation for the constancy of the speed of light \cite{landau2013classical} and finally, the link between the speed of light and the gravitational constant at the sight of the current understanding was unnatural, physically speaking.\\

Another weak point is the fact that the variation of the speed of light determines the properties of the gravitational field which in turn the gravitational field depends in the form how the speed of light changes in it, that is a circle thinking. For instance, the following equation in the modern physics is difficult to treat as field equation:

\begin{equation*}
\begin{aligned}
\laplacian{c} = \pdv[2]{c}{x} + \pdv[2]{c}{y} + \pdv[2]{c}{z} =0
\end{aligned}
\end{equation*}

As we know the energy conservation law is a statement concerning the independence with respect to time of a certain physical magnitude, therefore is natural in my opinion to encounter valid the conservation energy using the idea of the possible variation of the speed of light as Albert shown in his $1912$ article because there is no dependence with respect to time, which time?, is say; $\dv{C}{t}=0 \equiv \dot{C}=0$ as one can see. Therefore a lot of complications arises but despite of this I need to recognize that the idea is good in some aspect beyond that I don not think, at this understanding level, it will be necessary to deal with such touchy proposal idea.\\

\begin{center}
\textbf{Is it a good idea to establish a field theory upon the variation of one of the fundamental constants in Nature? \footnote{To consider other attempts concerning the variation of other universal constants like the gravitational constant see, for example, footnote $5$ }}
\end{center}

\textit{Does any physical idea has a true meaning or is just an intention to move from the non-location to a known one? Lost again $\cdots$}\\

The interval between two evens represent the $4-$distance in $\real^4$ and in turn is also the mathematical expression for the constancy of the speed of light with respect to the inertial reference systems, as I mentioned before. Therefore, the main idea is that $C$ is a constant physical quantity, in other words:
\begin{equation} \label{equ49}
\begin{aligned}
\mathrm{d}s^2=c^2 \mathrm{d}t^2-\mathrm{d} \va*{r}^2=c^2 \mathrm{d} t'^{2}-\mathrm{d} \va*{r}'^{2}= \mathrm{d} s'^{2}
\end{aligned}
\end{equation}

To see how useful is the equation [\ref{equ49}] , we need to see through it and make a link with the idea presented in the $1912$ article which, of course, is not easy to do. However it is possible to make an attempt in order to comprehend what Albert tried to communicate in using the possible variation of the speed of light. This was already done above and this makes it possible to conclude the main discussion concerning the Einstein's idea in the $1912$ year.

\epigraph{\vspace{-0.4 cm}
\textit{M. Abraham's idea and the physical mistake made in his article}}
{\textit{Again the consideration of the variation of the speed of light as suitable path to a new gravity theory}}
\vspace{-0.4 cm}
We have been studying two published articles concerning a plausible new theory of gravity one of them due to Albert Einstein in responds to the article due to M. Abraham but both contains basically the same physical idea, to change the speed of light, however, with a different point of view. If one stand on the basis ideas of these two articles we can notice that both solutions are wrong from the point of view of modern physics, one of the idea had as consequence the non-orthogonality condition deduced from the Minkowski's article \cite{lorentz1952principle} between the $4-$velocity and $4-$acceleration. I have been very careful in saying something concerning this issues of the $1912$ article but I still have the posture about the wrongness of both papers.\\

In the $1912$ article, Einstein made use of the invariance of the interval $\mathrm{d}s=\mathrm{d}s'$, no considering that this is valid in the case of the realm of the special theory of relativity and take into account the constancy of the speed of light. Therefore, there was not justification in doing use it to derive field equation which was really a poor mimic model of the Newton's celestial mechanics. I know that all this was written as good attempts but later on in $1916$ until maybe $1930$ Albert had in mind that in the presence of gravity a light-ray should undergo a variation of its speed. As we know, in both approach to the solution did not take into consideration the formalism that comes from the $1909$ paper by H. Minkowski, is say, all magnitude must to be $4-$dimension quantities for instance, force, acceleration and others. Here I am talking about the article wrote in $1912$ by Albert and try to address the mistake made to consider the variation of $C$ at least in the classical physical thinking.\\

Which is more subtle is the fact that the variation of the speed of light in turn leads us to the non-constant value of the gravitational constant. In my opinion, is no a good idea to propose a variation of the fundamental constant only for the reason we do not comprehend fully the meaning of it, so we would rather to try to find a cause of that, as I said there must be a geometrical reason of the constancy of the speed of light but is not so easy and that is the true research one must do. Einstein's article was a kind of attempt but nothing more than that, is not works as he expected but clean up the path for the final theory in which  he thought that the speed, really the velocity, has to undergo a change due to the curvature of space-time. But his theory of gravity as is very known has left far away Maxwell's field equations.\\

I don't have any good feeling nor a counter argument in using equation [\ref{equ49}] in order to derive a transformation which include a variation of the speed of light, even Albert, in the first and second order, he did not gave a good physical reason why the invariance of the interval hold true in the case he used for his theory. All the rest of the article are equations in the first approximation in contrast with the article \cite{martinez2022influence} where two simple statements gave us the same result with no approximation. In spite of all this that I have mentioned above, we can give more physical reason against the wrongness of the $1912$ article. For that, one must take in the scene the modern language to point out the mistake made in both article, by Albert and M. Abraham.\\

\textit{At least something have to be true, either the speed of light is not so fundamental or the main property known is really what it is $\cdots$}\\

Let's now discuss the M. Abraham $1912$ paper concerning what he called a new theory of gravity where he took the idea of the variation of the speed of light as sound argument in order to establish a suitable gravity theory.

\section*{\S15.\, M. Abraham's new theory of gravity, is everything boil down to the variation of the speed of light, without sound real physical argument?}

M. Abraham based his idea of his new theory of gravity on the Minkowski's paper of $1909$, \cite{lorentz1952principle} and particular the idea of orthogonality condition proposed by Herman was taken by Abraham as not valid when one pass from the special theory, no presence of gravity, to non-inertial reference frame, namely a particular region where is a presence of a gravitational field. Therefore, it is important at this point to make some remarks about Minkowski's paper and then list the error made in the paper by M. Abraham.\\

The orthogonality condition is related with both $4-$quantities, velocity and acceleration. This can be written as follows:
\begin{equation} \label{equ50}
\begin{aligned}
u_\alpha w^\alpha & =0, \\~ \dot{c} & =0\\
u_\alpha  u^\alpha & = c^2 \Leftrightarrow  \dv{u_{\alpha}}{s} u^\alpha+u_\alpha \dv{u^\alpha}{s} & = 2c\dot{c} \dfrac{\mathrm{d}\tau}{\mathrm{d}s}
\end{aligned}
\end{equation}

So, as the dot product is a symmetric mathematical operation, even in the $\real^4$-dimensional space-time, we can sum up the last two terms of equation [\ref{equ50}] and taking into consideration the fact that the speed of light is a constant physical magnitude, one can obtain:

\begin{equation*}
\begin{aligned}
u_\alpha \dv{u^\alpha}{s}=0 \Leftrightarrow u_\alpha w^\alpha=0
\end{aligned}
\end{equation*}

What were made by Abraham in his paper were to consider the speed of light as a function of the proper time but without a physical justification. When really is not like so, in the modern language the orthogonality condition state that both $4-$vectors are perpendicular and to go to the case of a presence of gravity one just need to replace the ordinary derivative by the covariant derivative, is say, $\dd \rightarrow \partial_{;\alpha}$.
Taking seriously the idea, physicist had to take the variation of the speed as physical phenomenon even when was necessity only to compute the angle deviation and all the other physical consequences in the $1911$ paper were derived without it. However, the main question is still persist in mind, at least to me that question should be answer, why Einstein after his revolutionary theory persisted in this idea of possible variation of the speed of light?\\

What was done in the article \cite{martinez2022influence} has more physical sense because I did not use the conjecture of such a variation and all the physical results were the same and also the computation of the angle deviation gave the same result and that was something I did not expect at all, so one can conclude that something is missing in the idea or in the calculation, because what really I expected to find was the value given by the general theory of relativity; is say, twice of the calculated. Therefore, it is necessary to revise that particular point. But now, let's continue with Abraham's idea.\\

Now, the variation seems to be with respect to the proper time $\tau$, the variation of the speed of light, and not with respect the position at some point in $\real^4$ as one could verify in the article \cite{abraham2007new}. in turn of that one could deduce from the modification made by Abraham concerning the orthogonality condition that $\dot{c}= \frac{\mathrm{d}c}{\mathrm{d}\tau}=0$ so the equation:
\begin{equation} \label{equ51}
\begin{aligned}
mc \dfrac{\mathrm{d}(ck^{-1})}{\mathrm{d}\tau
}=0
\end{aligned}
\end{equation}
make no sense at all due to the fact that if there is no proper-time dependence there is no variation in the speed of light, otherwise if there is a proper-time dependence all the approximation made in the article wrote by M. Abraham will have an intrinsic contradiction as we pointed out using equations [\ref{equ55}] and [\ref{equ44}]. In the article mentioned there were used two approximations which yields to the energy conservation law in the case of low velocities, low with respect to the speed of light,  but the two approximations have a crucial problem, they are not consistent mathematically speaking. Another point to emphasize the mistake made could be about the intensity of the gravitational potential which has a relation with the speed of light, is say, if the gravitational potential will become infinity at some finite region the speed of light could has the same behavior and that is a serious physical problem.\\

In the modern language, one could say, if the gravitational potential will become infinity light would not be capable to escape from that region due to the finite and constant value of its speed which sounds like a more physical idea than the previous one. Of course, one should be able to understand that all these statements have a local validity and the idea is to study the mistake made in the year of $1912$ concerning the attempt for a new theory of gravity which is a little bit complicated to do. Therefore, in my opinion is very natural to kept constant the speed of light even in the presence of a gravitational field and try to justify this assertion with some geometrical idea and in turn may gives us a geometrical interpretation of the electromagnetic field as Albert Einstein pointed out many times during his last active years of scientific investigations.\\

Hence, the deviation of light from its rectilinear path is a consequence of its finite speed so there is not need of the equivalence principle, apparently, one could say that the equivalence principle is a way to interpret all the phenomena occurring in a non-inertial reference system as phenomena occurring in a inertial reference system with a particular gravitational field. Moreover, if there is really an influence of the curvature to the motion of light Maxwell's equations should have to say something concerning that but is not so and that is a big problem because there is no place to Maxwell's field equations in the geometrical interpretation of gravity since its formulation by Albert Einstein.\\

One can not go further assuming something without experimental aids that support the idea, in this case the variation of the speed of light, however from the historical point of view is educational to know how difficult was to obtain an useful equations that describes the behavior of gravity. Might be possible, of course, that this phenomenon will be useful in the far future when we left behind the basis physics that is known and take seriously a possible unification between gravitational and electromagnetic fields with quantum mechanics, but will be necessary a geometrical interpretation of the phenomenon, Why $C$ have to be constant classically speaking?\\

\textit{What  was wrong with M. Abraham's idea?}$\cdots$\\

Everything boils down to the orthogonality condition in the sense that this condition represents the magnitude of the $4-$velocity and must be invariant under general transformation and besides that, must be a constant physical magnitude. So, from the point of view of the modern physics it has not physical meaning apart from the one we could assign to it.

\epigraph{
\textit{Kept the speed of light as a constant is a physical fundamental principle even must be true in a presence of a gravitational field}}
{\textit{A scalar term of interaction and the energy conservation law}}
..........
\section*{\S$16$. \, Weyl's idea concerning unification fields between gravity and electromagnetism\\ \vspace{-1 cm} \center{Are we in a big trouble?}}

Weyl proposed his idea about unification fields between gravity, as Einstein formulated it in $1916$, and the Maxwell's field equations. The start point of Weyl's idea is a generalization of the Riemann's geometry but in this case with an additional property concerning a variation in the length of vectors when are moving  from one point to another in a differential manifold, is say: 
\begin{equation} \label{equ52}
\begin{aligned}
\dd A=-A\Psi_{\alpha}\dd x^\alpha
\end{aligned}
\end{equation}

From equation [\ref{equ52}] Weyl was capable to derive a new Christoffel connection for the proposed geometry but later on it was rejected by Albert using physical arguments, see \cite{weyl1952gravitation}, in spite of that the idea really unify both fields and  it was took for the quantum mechanics realm and is nowadays an useful tool to study some symmetry aspects. Therefore, the idea was abounded in the geometrical sense as a plausible and sound solution to the unification problem. Here there is a question, How weyl's idea enters to this philosophical discussion? As I mentioned in page $6$ equation $10$, perhaps the most simple proposed for the variation of the speed of light could be equation [\ref{equ47}] however all the physical complications that will appear is enormous as Einstein pointed out in order to invalid the Weyl's idea.\\

Any advance in physics always has something weird and useful, the point is to see that and be able to develop the idea until its final physical consequences as usually is. Hence, one can propose a new idea but could be wrong and that's okay because that means to cancel out one wrong way. So, instead of take seriously the variation of the speed of light one could take another physical magnitude,for instance, the rest mass of a point-like particle, or a system, and use the Weyl's idea to see what we can get from it. In sight of this we can do that, so that will be the main course in this last topic of this little investigation.\\

Consider a point-like particle which is moving in a presence of a $4-$dimensional field, $\Psi_\alpha$, so instead of the variation of the length of an arbitrary vector we are going to take a possible variation of the rest mass of the particle, or in a more extend way could be a system, mentioned above in the following way: 
\begin{equation}\label{equ53}
\begin{aligned}
\dd m_0 = -m_0 \Psi_\alpha \dd x^{\alpha} \Leftrightarrow \dd m_0 =-m_0\Psi_\alpha v^\alpha \dd s\\
\left| v^\alpha \right |=1\\
\left \{
\begin{aligned}
\Psi\alpha v^\alpha & = k =\left \{ \begin {aligned} k = \text{constant}\\ k= k(s^-1)  \end{aligned} \right.\\
\dd m_0 & = - m_0 k \dd s
\end{aligned}
\right.
\end{aligned}
\end{equation}

Considering the constrain condition imposed to equation [\ref{equ53}] one can obtain an interesting equation of motion, in the sense of the geodesic equation like a kind of generalization. Doing that one can get:

\begin{equation} \label{equ54}
\begin{aligned}
\Psi_{\alpha} v^\alpha = k {v_\alpha} v^\alpha \Leftrightarrow \Psi_\alpha v^\alpha-\Psi_\gamma v^\gamma v_\alpha v^\alpha=0
\end{aligned}
\end{equation}

Equation [\ref{equ54}] could be written as follows, taking into consideration the orthogonality condition between both $4-$dimensional quantities, velocity and acceleration:

\begin{equation} \label{equ55}
\begin{aligned}
v^\alpha \left(\Psi_\alpha \eta_{\gamma \theta} v^\theta v^\gamma-\Psi_\gamma \eta_{\alpha \theta} v^\theta v^\gamma \right)=0 \Leftrightarrow \dv [2]{v_\alpha} {s}+ \dfrac{1}{2} \left(\Psi_\gamma \eta_{\alpha \theta} + \Psi_\theta \eta_{\alpha \gamma}-2 \Psi_\alpha \eta_{\gamma \theta} \right) v^\theta v^\gamma=0
\end{aligned}
\end{equation}

So that equation [\ref{equ55}] represents the trajectory in absence of any gravitational field, on the contrary when there is a gravitational field one can simple replace the ordinary derivatives by covariant derivatives. Therefore, in this manner with all the computations done we have obtained the modification geodesic as Weyl got with some slightly difference, the equation differs by the factor of $2$ in the third term of the second equation in [\ref{equ55}]. The physical meaning of equation [\ref{equ55}] remains unclear up to now but some near application in the realm of cosmology should be possible if one stop for a moment to think about it. Let's now write in a different way the third equation in [\ref{equ53}]:
\begin{equation} \label{equ56}
\begin{aligned}
\dd m_0 & = -m_0 \mu \dd \tau \\
\text{Where:} \\
k & = \dfrac{\mu}{c}\\
-\dfrac{\dd m_0}{m_0 \dd \tau} & = \mu
\end{aligned}
\end{equation}

In this way we were capable to get an equation on the variation of the rest mass of a particle which undergo the influence of the $4-$dimensional field, $\Psi_\alpha$. Being $\mu$ a positive, negative or constant value parameter, what is not so plain at this point is the physical meaning of all this. If one integrate equation [\ref{equ56}] we obtain a relationship between the value of the rest mass at some point in $\real^4$ and the proper time along a world line element, is say:
\begin{equation} \label{equ57}
\begin{aligned}
m_0 \left(2 \right)=m_0 \left(1 \right) exp\left(-\int \limits_{1}^2\mu \sqrt{1-\dfrac{v^2}{c^2}}dt\right)
\end{aligned}
\end{equation}

Right from equation [\ref{equ57}] one could obtain for a light ray both values of the rest mass from one point to another and turns out to has the same value: $ m_0 \left(2 \right) = m_0 \left(1 \right)= \text{constant}$, and particular we can set $m_0 \left(2 \right)= \cdots m_0 \left(1 \right) =0$ for light but the zero rest mass for light was not deduced from equation [\ref{equ57}] is a value one could assign because is what it is known from physical experiments.\\

If one look at the equation [\ref{equ57}] it is possible to deduce from it that; using the equivalence principle, the components of the metric tensor could be seen  as some kind of random variables,therefore all the components can undergo a fluctuation in its value which in turn means that in small step the trajectory of a particular body should experience a deviation with respect to its original trajectory which happens with the most inner planet of the solar system, Mercury, of course this is only just an idea. This, of course, have a lot of implications in the geometrical sense but is not so easy to give a physical interpretation in order to use it to find solutions to the problems of cosmology yet. In other words, for instance, the $g_{00}$ of the metric have to satisfy the inequality condition \[0 < 1-\dfrac{\phi}{c^2} \leqslant 1 \] for all possible values assigned to the metric, which means that some geometry configuration would not be possible.\\

Return to the idea of the change of the rest mass one could write the following equations related with the idea mentioned in last paragraph,for instance, equation [\ref{equ20}] could be written as follows: 
\begin{equation} \label{equ58}
\begin{aligned}
-\dfrac{\mathrm{d}m_0}{m_0 dt} &= \mu \sqrt{1-\dfrac{v^2}{c^2}}\\ \text{If one consider the equivalence principle, we have:}\\
v^2 &= 2 \, \text{acceleration}\,  x \,\rightarrow 2 \phi \\
\mu \sqrt{1-\dfrac{v^2}{c^2}} \rightarrow \mu \sqrt{1-2\dfrac{\phi}{c^2}}  & \approx \mu \left(1-  \dfrac{\phi}{c^2}\right)
\end{aligned}
\end{equation}

Clearly one can see that the temporal component of the metric in first order approximation behave as random variables, could undergo a fluctuation, as I mentioned above many times, in its values and that is something completely new because makes me think in another way what possible be the space. I would like to give more deeply arguments concerning this point of view as new way of interpretation of the components of the metric tensor, which one could think would be the cause of the non-closed orbit of the planet mercury. In other words due to the fluctuation of the components of the metric in each small step, the trajectory will deviate from its original path so in the end the trajectory is not what we can expected classically, is say, some probabilistic influence is there.\\

Moreover, a radioactive elements could have a different mean life decay due to the location where is in space-time which means basically the influence due to the gravitational potential, $phi$. Which seems to me very interesting; perhaps this is a possible path to the unification  with quantum mechanics but in a different way, but is just only an idea without a mathematical support. I would like to address the attention to this and be capable to cover this topic in another article.\\

For example if one put oneself in a particular reference system and set the condition that the  decay-life have to be the same for this observer we can obtain physical condition related with it and the potential gravitational play a crucial role here, this computation is easy to do and is not necessary to do it here.\\

\textit{Weyl's idea of unification fields $\cdots$, was A. Einstein wright about his comments concerning the idea?}\\

Riemann's idea concerning of the background geometry used in solve the gravitational problem has the property to keeping invariant the length of any $4-$dimensional vectors, which means that the convariant derivative of the metric tensor is zero and the symmetry property of the Christoffel symbols of the second kind, is say: 
\begin{equation}\label{equ59}
\begin{aligned}
g_{{\mu \nu};\alpha} & =0\\ \Leftrightarrow \\
\Gamma_{\mu \nu}^\alpha-\Gamma_{\nu \mu}^\alpha & =0
\end{aligned}
\end{equation}

As we know the metric tensor is a form to measure distances and other properties in Riemann's geometry, therefore if there is a variation in the length of the vectors the covariant derivative of the metric will not be zero any more. In the Weyl's article \cite{weyl1952gravitation} one can found a derivation of the geodesic but there is no mention about the zero derivative of the metric. However, with some mathematical manipulations is possible to find the equivalence mentioned above. Let's try it out, we are going to use one of the equation that appears in the article and find the equivalence.
\begin{quotation}
$\cdots$ If ti is known, then the quantities $\Gamma$ are determined by equation $(6)$ or \footnote{see the article mentioned \cite{weyl1952gravitation}}
\begin{equation*}
\Gamma_{i,kr} + \Gamma_{k,ir}= \pdv{g_{rk}}{x_r}-g_{ik} \psi_r
\end{equation*}

and the symmetry property $(5)$. \texttt{The metrical connection of the space depends not only on the quadratic from $(2)$ $\cdots$ but on the linear form $(7)$}
\end{quotation} 
\vspace{-2 pt}

Using the equation from Weyl's article wrote inside the quotation we can rewrite it in the following form:

\begin{equation} \label{equ60}
\begin{aligned}
g_{\alpha \gamma ; \theta} &= g_{\alpha \gamma ; \theta}-g_{\alpha n} \Gamma_{\gamma \theta}^n-g_{n \gamma} \Gamma_{\alpha \theta}^n \\\Leftrightarrow \\ g_{\alpha \gamma;\theta}& =g_{\alpha \gamma} \psi_\theta \\
g_{\alpha \gamma} &= g_{\gamma \alpha} \\ \text{and also}\\
\Gamma_{\alpha \gamma}^\theta &= \Gamma_{\gamma \alpha}^\theta
\end{aligned}
\end{equation}\\

Considering the symmetric properties of both tensors, the metric and the Christoffel second symbols. It is interesting to remember that the anti-symmetric property of the metric tensor was extensively studied by Albert Einstein in his own way to try to formulate a unification theory. The $4$-dimensional field was treated by Weyl as the electromagnetic part in the generalization geometry proposed by himself, even though the idea really unify both fields it has a weak point and this was that there are no physical arguments in order to introduce the variation of the length of vectors. In other words, there was not reason at all why this variation of the length has to be associated with the interaction or part of the consequence  concerning the presence of the EM field in a region with gravitational field, is say, how gravity affects the EM field and vice versa.\\

Something very similar with both ideas came from Albert and Abraham concerning the absence of sound physical arguments in order to make the conjecture of the variation of the speed of light as tenable physical theory; Seems to work at some point but it is not possible to deduce physical phenomena to face with experiments and that is a weakness of the idea even whether in the unification this constancy could be not valid any more, besides, the main thing is, Why?.\\

\textit{Albert Einstein and the total rejection of the Weyl's idea of the unification fields $\cdots$}\\

Most of the time when new ideas is present to the scientific community two things might happens, one is to pay a lots attention and see if the idea should be right or not via test experiments and the other one is to reject it but here is where something take place; Others take the idea, out of the realm  in which the idea was originally created, and found a place for it, for instance, like in the realm of the quantum field theory. That is what happened to Weyl's idea, however, why Albert was so rigid with this idea? The problem, it was a touchy thing because Einstein's idea concerning relativity do not match in any sense with the consequences of the Wey's idea, of course for Einstein. But it was not a problem to Einstein the possible variation of the speed of light where he was the first to use the idea so a question arises here, why?\\

In the article \cite{weyl1952gravitation} at the end of the paper one can read: 
\begin{quotation}
$\cdots$ Such a definition of the line element $\mathrm{d}s$ would become illusory only if the assumption concerning standard lengths and standard clocks was not valid in principle; this would be the case if the length of standard rod depend on its history $\cdots$
\end{quotation}

One can avoid the question concerning if the variation of the rest mass, as was proposed here, could violate the energy conservation law; however, put it this apart from that makes it a lot of sense the idea if one is willing to push it and see what the idea is try to address. Perhaps in another article should be a good idea to cover this issue and try to put light on it. The idea is to point out that all the components of the metric could have a different physical interpretation, in the sense that I expressed  above, so in this manner should be possible to be able to know what kinds of geometries are allows and which not.
....
\section*{\S$16.1$. Closing Remarks of Th Second and Last Part}
 \begin{enumerate}
 \item [$\bullet$] It is crucial to understand the importance concerning the role play by the speed of light in both theories, the special and the general theory of relativity, at least to keep constant the speed until appears a sound reason to make such a modification in the foundation of the modern physics, perhaps in the process of unification between the most two understandable fields.\\
 
 There is no reason or sensible one to make such a radical conjecture now, I think we need to kept that constant, the speed of light, and to work on that foundation in order to comprehend what we are not capable to understand nowadays.
 
 \item [$\bullet$]The deviation of the light ray near to a strong gravitational field, as weak ones, could be seen as interaction process between both physical fields which is  completely different from A. Einstein's point of view but possible. Therefore, a more deeply study of the basic interaction term should have to be done in relation to the geometrical process made for the gravitational field by A. Einstein.
 \begin{equation*}
 \begin{aligned}
 \text{Basic Interaction Term} \rightarrow \hbar \dfrac{\phi}{c^2}\\
 c & = \cdots \text{constant}
 \end{aligned}
 \end{equation*}
 
 It is also a good to kept constant the speed of light in a presence of any gravitational field, moreover it is important to seek a geometrical interpretation of it, so if there is a necessity to make a conjecture of such a variation it has to do with a possible unification of both fields which are  the most known in the physics realm, as I mentioned many times, at least in first order solution  to the problem or maybe not. \\
 
 \item[$\bullet$] Even though the conjecture of a variation of the speed of light permit a way to obtain a gravitational equation it has not a real physical meaning in the sense that with that idea the equation get is not useful in any physical situations. Hence, both ideas from Albert and Abraham were only an approach to the surface of what later on Albert did. If Albert persisted in this possible variation were not clear the reasons behind that, but all possible phenomena could be explain more naturally with the postulate concerning the constancy of the speed of light.\\
 
 \item [$\bullet$] An approachable solution to the unification field problem could be possible if we have a geometrical interpretation of the electromagnetic phenomenon as Weyl tried to do in his paper \cite{weyl1952gravitatio}. It might sounds difficult at this moment but it is necessary to clarify the reason why up to now is no possible a first order unification, at least, to the unification field problem, to say this I'm based my opinion on the tangible fact that gravity and electromagnetism are both the only two fields that we have a direct interactions and one of them is a friendly  field in the sense  that in some form one could manipulate it very easily, for instance, it is possible to construct particle accelerator and is also possible to launch a rocket from the Earth's surface. So, here is a big problem with more questions than answers:
 \[  G_{\mu \nu}= \alpha T_{\mu \nu} \, \cup \, F_{\mu \nu}\] 
 
 \item [$\bullet$]One interesting issue concerning this situation is that the field equations are valid locally and in order to make any statement we need to comprehend how to use the metric tensor in the study of large geometrical structure, cosmological speaking, perhaps this is a serious problem that one have to immediately pay attention and in turn to find a tenable and sound solution but only the near future could say something about it.
 \end{enumerate}
 
\section*{\S$16.2$. Conclusion}

 I still have a long way to go to understand the most basic thing like the concept of space, if there is one, as well time. But to think different is the key in the scientific investigation, in the first article the proposed made concerning a process of interaction between both fields was a good idea and I was able to get the same results, however, there are more to do there. Now with this article I realized that there is no good idea in order to solve the unification problem and most of the physics is working on the same basis which others left behind, like Klein's ideas of unification, for instance, so a little step was made from $1926$ up to the present day.\\
 
 In my opinion, I would say that, we need to seek a geometrical interpretation of the EM and try to reconcile this field with its better friend \textit{GRAVITY}, if one can do this step, the solution will be closer than we could think but for sure will give us more question than answers. The Nature is an intelligence organism. Therefore,we can not continue walking in the wrong way, what we need is an idea that possible will change the physics from its basis if there are ones.\\
 
 All the attempts made from the geometrical formulation of gravity to unify it with other fields is still unfruitful; perhaps the quantum mechanics ideas could be a way but it is still a premature to give a final word concerning that but there is no doubt that have to be in account, now the question is, in what form will enter in order to solve this puzzle? behold, the question to answer.

\section*{\S17. Acknowledgment}

The author would like to thank all the comments and suggestions made by close friends and their moral support in the process of write the article which is always a hard task to do. I also would like to dedicate this tiny contribution on the philosophy of physics a friend of mine, \textit{Yerson Suarez} who passed away but his presence is still with me looking at the world in a different way.\\

And I also would like to express my great appreciation a friends of mine, \textit{Roberto Pomares and Christopher Herrera}, whose dedicated their spare time to discuss some physical ideas and long discusses concerning the main subject of this article and for their support and friendship as well. I appreciate also the help during this difficult time  from a family members whose moral support was invaluable, \textit{MSc. Guillermo Martinez}. \\
\begin{tikzpicture}
\draw (-2,0) -- (10,0);
\end{tikzpicture}

My gratitude and appreciation to the being who made this possible: \\

\textbf{To God, who can make everything possible!}

\begin{tikzpicture}
\draw (-2,0) -- (7,0);
\end{tikzpicture}

\pagebreak

\section*{\S $18.$ \,Final Comments About this Little Research}

In $1911$ Albert Einstein took the advantage concerning the development of the new theory of gravity however he not realized all the useful mathematical tools in his time like the unification of the space-time due to H. Minkoswi. Apart from that his idea of possible variation of the speed of light was the only light through many years of seeking until he arrived with the right solution, he finally realized that  was necessary to take all the components of the metric tensor into account, \( g_{\nu \mu }\). And wrote a several papers concerning the meaning of this geometrical quantity, however Riemann in the early development of his geometry pointed out the same ideas as Einstein found in $1916$ approximately.\\[3 pt]

So it was not totally new but what was new is the fact that Einstein was able to formulate a gravity theory and not only to point out the possibility. Even his idea of variation of the speed of light was a possible solution for him in those years $1907-1912$, and the final result of the theory showed up that was not necessary and perhaps unnatural to suggest that, but the question remains until now, is say, Why does exists such limit in the information transmission  in Nature?The main purpose and idea to write this was not to contradict anyone whom are expertise in the field of gravity but rather to point out that all the problems nowadays physics is facing is due that there's not a clearly understanding  about what's really is space and time as well. \\

It is necessary new ideas and therefore new approach to the solution, so I'd like to try to find some different approaches to the solutions and I'll be eager to prove it whether I'm wrong or not. Please any comments about this research and the proposed idea let me know it as soon as possible.
\vspace{2 cm}
\begin{center}
\begin{tikzpicture}
\draw (2,0) -- (9,0);
\end{tikzpicture}
\end{center}
\begin{center}
Submitted date: (October $8$, $2022$)\\
 Received on the $28^{\text{th}}$ of September $2022$\\ (\textbf{\,Version Number $6$, $2023$} )\\
\end{center}
\begin{center}
\begin{tikzpicture}
\draw (2,0) -- (14,0);
\end{tikzpicture}
\end{center}
\begin{center}
Submitted date of the second and last part: (May $19$, $2023$)\\
Received on the $15^{\text{th}}$ of May $2023$\\
(\textbf{Version Number $1$, $2023$)}\\
\end{center}
\begin{center}
\begin{tikzpicture}
\draw (2,0) -- (14,0);
\end{tikzpicture}
\end{center}
\begin{flushright}
Managua, Nicaragua(May $19$, $2023$)\\
\copyright For Academic Use Only\\
\href{emailto: rigoberto3martinez@gmail.com}{\mbox{\texttt{rigoberto3martinez@gmail.com }}}\\
\href{emailto: zeit71989@outlook.com}{\mbox{\texttt{zeit71989@outlook.com }}}
\end{flushright}
\pagebreak

\bibliographystyle{plain}
\bibliography{FirstArticle}
\end{document}